\documentclass[a4paper,UKenglish,cleveref, thm-restate]{lipics-v2021}
\usepackage{quoting}
\usepackage{environ}
\usepackage{xcolor}
\usepackage{array}
\usepackage{tikz}
\usepackage{colortbl}
\usepackage{multirow}
\usepackage{todonotes}
\usepackage[appendix=inline]{apxproof}
\usepackage{subcaption}
\usepackage[labelsep=quad, skip=3pt]{caption}
\usepackage{graphicx}

\usetikzlibrary{calc,shapes,patterns,positioning,fit,backgrounds,decorations,chains}
\usetikzlibrary{decorations.pathmorphing, shapes}
\usetikzlibrary{arrows}
\usetikzlibrary{decorations.markings}
\usetikzlibrary{arrows.meta}
\usetikzlibrary{chains,fit,shapes, shapes.multipart}
\usetikzlibrary{decorations.pathreplacing}

\usepackage[french,linesnumbered,lined,boxed,ruled,noend]{algorithm2e}

\usepackage{amsmath,amssymb,amsfonts}
\usepackage{amsthm,nicefrac}
\crefname{subsection}{subsection}{subsections}

\pgfdeclarelayer{bg}    % declare background layer
\pgfsetlayers{bg,main}

\usepackage{ulem}
 
 \usepackage{paralist} 
 \usepackage[shortlabels]{enumitem}
\usepackage{float}
 \usepackage{amssymb}
\newlist{todolist}{itemize}{2}
\setlist[todolist]{label=$\square$}
\usepackage{pifont}

\usepackage{xifthen,ifthen}

\newcommand{\newJ}[2][]{{\ifthenelse{\isempty{#1}}{}{\todo[color=orange!50]{#1}}\color{orange}#2}}
\title{Degreewidth: a New Parameter for Solving Problems on Tournaments} %TODO Please add

\author{Tom Davot}{Université de Technologie de Compiègne, CNRS, Heudiasyc (Heuristics and Diagnosis of Complex Systems), CS 60319 - 60203 Compiègne Cedex, France}{tom.davot@hds.utc.fr}{https://orcid.org/0000-0003-4203-5140}{This work was carried out and funded in the framework of the Labex MS2T. It was supported by the French Government, through the program ``Investments for the future'' managed by the National Agency for Research (Reference ANR-11-IDEX-0004-02)}%TODO mandatory, please use full name; only 1 author per \author macro; first two parameters are mandatory, other parameters can be empty. Please provide at least the name of the affiliation and the country. The full address is optional. Use additional curly braces to indicate the correct name splitting when the last name consists of multiple name parts.

\author{Lucas Isenmann}{Université Paul Valéry, Montpellier III, Montpellier, France}{lucas.isenmann@laposte.net}{ https://orcid.org/
0000-0002-1460-269X}{}

\author{Sanjukta Roy}{Faculty of Information Technology, Czech Technical University in Prague, Prague, Czech Republic}{sanjukta.roy@fit.cvut.cz}{https://orcid.org/0000-0003-3633-542X}{}

\author{Jocelyn Thiebaut}{Faculty of Information Technology, Czech Technical University in Prague, Prague, Czech Republic}{jocelyn.thiebaut@cvut.cz}{https://orcid.org/
	0000-0002-4550-8399}{Supported by the CTU Global postdoc fellowship program}

\authorrunning{T. Davot, L. Isenmann, S. Roy, and J. Thiebaut} %TODO mandatory. First: Use abbreviated first/middle names. Second (only in severe cases): Use first author plus 'et al.'

\Copyright{Tom Davot, Lucas Isenmann, Sanjukta Roy, and Jocelyn Thiebaut} %TODO mandatory, please use full first names. LIPIcs license is "CC-BY";  http://creativecommons.org/licenses/by/3.0/


 %TODO mandatory; please add comma-separated list of keywords

\category{} %optional, e.g. invited paper

\relatedversion{} %optional, e.g. full version hosted on arXiv, HAL, or other respository/website
\acknowledgements{We would like to thank Frédéric Havet for pointing us a counter-example to the polynomial running-time algorithm to decide if a tournament is sparse.} %; it lead us to the direction of tournaments $U_n$.}%optional

\nolinenumbers %uncomment to disable line numbering

\EventEditors{John Q. Open and Joan R. Access}
\EventNoEds{2}
\EventLongTitle{42nd Conference on Very Important Topics (CVIT 2016)}
\EventShortTitle{CVIT 2016}
\EventAcronym{CVIT}
\EventYear{2016}
\EventDate{December 24--27, 2016}
\EventLocation{Little Whinging, United Kingdom}
\EventLogo{}
\SeriesVolume{42}
\ArticleNo{23}
\tikzstyle{smallvertex}=[circle, draw, fill=white, inner sep=2pt]
\tikzstyle{m}=[line width=1mm]

\definecolor{goldenpoppy}{rgb}{0.99, 0.76, 0.0}
\definecolor{iris}{rgb}{0.35, 0.31, 0.81}
\definecolor{forestgreen}{rgb}{0.13, 0.55, 0.13}

\newcommand{\BTSAT}{\textsc{\mbox{Balanced 3-SAT(4)}}\xspace}

\newcommand{\FAST}{\textsc{\mbox{FAST}}\xspace}

\newcommand{\myparagraph}[1]{\noindent\textbf{#1}}
\newtheorem{construction}{Construction}

\makeatletter
\@ifpackagewith{apxproof}{appendix=inline}{}{}
\ifthenelse{\equal{\axp@appendix}{strip}}{}{}
\NewEnviron{sketch}
{ \ifthenelse{\equal{\axp@appendix}{inline}}{}{
    \begin{proofsketch}
      \BODY
    \end{proofsketch}
  }}
\makeatother

\newtoggle{teach}
\toggletrue{teach}
\togglefalse{teach}
\NewEnviron{teachr}
{\iftoggle{teach}{\BODY}{}}

\newtheoremrep{theorem}[ctheorem]{Theorem}

\newtheoremrep{lemma}[clemma]{Lemma}

\newtheoremrep{corollary}[ccorollary]{Corollary}

\newtheoremrep{observation}[cobservation]{Observation}
\newcommand{\infor}[1][\sigma]{\ensuremath{\prec_{#1}}}
\newcommand{\de}[1][\sigma]{\ensuremath{d_{#1}}}
\newcommand{\dw}[1][\sigma]{\ensuremath{\Delta_{#1}}}
\newcommand{\DW}{\ensuremath{\Delta}}

\begin{document}

%\begin{itemize}
%	\item TO DO LIST:
%	      \begin{todolist}
%		      \item[\done] Change title -> T: Ok
%		      \item Recheck Abstract: S and T ????
%		      \item recheck Introduction: T(Ok)
%		      \item[\done] rewrite poly sparse and remove some of the pseudocode:  L
%		      \item sketch proofs
%		      \begin{todolist}
%			      \item[\done] regular tournaments : J \textcolor{red}{$\rightarrow$ TO CHECK}, L: ok, added smthg
%			      \item 3-approx : L \textcolor{red}{$\rightarrow$ TO CHECK}
%			      \item[\done] U-tournaments 2 sparse patterns : J  \textcolor{red}{$\rightarrow$ TO CHECK}
%			      \item[\done] DS FPT : S \textcolor{red}{$\rightarrow$ TO CHECK}
%			      \item[\done] FAST poly : J \textcolor{red}{$\rightarrow$ TO CHECK}
%			      \item[\done] FVST NPH : T
%		      \end{todolist}
%		      \item  Recheck conclusion
%		      \item 12 pages : ``Authors are invited to submit an extended abstract or full paper with at most 12 pages (excluding the title page and the references section) to the appropriate track. The title page consists of the title of the paper and the abstract, but *no* author information. The first section of the paper should start on the next page.''
%	      \end{todolist}
%\end{itemize}

\maketitle

\begin{abstract}

	In the paper, we define a new parameter for tournaments called degreewidth which can be seen as a measure of how far is the tournament from being acyclic. The degreewidth of a tournament $T$ denoted by $\DW(T)$ is the minimum value $k$ for which we can find an ordering $\langle v_1, \dots, v_n \rangle$ of the vertices of $T$ such that every vertex is incident to at most $k$ backward arcs (\textit{i.e.} an arc $(v_i,v_j)$ such that $j<i$). Thus, a tournament is acyclic if and only if its degreewidth is zero. 
	Additionally, the class of sparse tournaments defined by Bessy \textit{et al.} [ESA 2017] is exactly the class of tournaments with degreewidth one.
	
	We first study computational complexity of finding degreewidth. Namely, we show it is NP-hard and complement this result with a $3$-approximation algorithm. We also provide a cubic algorithm to decide if a tournament is sparse. 
	
	Finally, we study classical graph problems \textsc{Dominating Set} and \textsc{Feedback Vertex Set} parameterized by degreewidth. We show the former is fixed parameter tractable whereas the latter is NP-hard on sparse tournaments.  Additionally, we study \textsc{Feedback Arc Set} on sparse tournaments.
	
\end{abstract}

\newpage

\section{Introduction}
Tournaments form a very rich subclass of digraphs which has been widely studied both from structural and algorithmic point of view~\cite{BJG08}.
They correspond to the class of directed graphs (digraphs) for which there is exactly one arc between each pair of vertices.
Unlike complete graphs, a number of classical problems remains difficult in tournaments and therefore interesting to study. These problems include \textsc{Dominating Set}, \textsc{Winner Determination}, or maximum cycle packing problems. For example, \textsc{Dominating Set} is W[2]-hard on tournaments with respect to solution size~\cite{downey1995parameterized}.
However, many of these problems become easy on acyclic tournaments (\textit{i.e.} without directed cycle).
Therefore, a natural question that arises is whether these problems are easy to solve on tournaments that are close to being acyclic. Several parameters can be considered in this regard.
% fas
For example, a \textit{feedback arc set} (fas) is a collection of arcs that, when removed from digraph (or, equivalently, reversed) makes it acyclic. Hence, it is common to consider the size of the minimum fas as a measure of the distance to acyclicity.
This parameter has been widely studied, for numerous applications in many fields, such as circuit design~\cite{Johnson75}, or artificial intelligence~\cite{BGNR98, Dechter90}. More specifically on tournaments (the problem is then called \textit{FAST} for {\sc Feedback Arc Set in Tournaments}), the question whether it is difficult to compute a minimum-FAS on this class of digraphs remained opened for over a decade before being proven NP-complete~\cite{Alon06, CTY07}.
%were published~\cite{Alon06, CTY07}.
%Conjectured NP-complete by Bang-Jensen and Thomassen~\cite{BT92} in 1992, Ailon \textit{et al.}~\cite{ACN08} were the first ones to its NP-hardness under randomized reductions.
%Since then, deterministic reductions were published~\cite{Alon06, CTY07}.
From the approximability point of view, van Zuylen and Williamson~\cite{ZW09} provided a 2-approximation of FAST, and Kenyon-Mathieu and Schudy~\cite{KS07} a PTAS algorithm.
On the parameterized-complexity side, Feige~\cite{Feige09} as well as Karpinski and Schudy~\cite{KS10} independently proved a $2^{O(\sqrt{k})} + n^{O(1)}$ running-time algorithm.
Another way to define FAST is to consider the problem of finding an ordering  of the vertices $\langle v_1, \dots, v_n \rangle$ minimizing the number of arcs $(v_i, v_j)$ with $j<i$; such arcs are called \textit{backward arcs}. 
%total strict order $\prec$ of the vertices minimizing the number of arcs $(a,b)$ such that $b \prec a$; such arcs are called \textit{backward arcs}. 
%When studying a tournament $T$, it is common to equip it with an ordering of its vertices (defined formally later).
Then, it is easy to see that a tournament is acyclic if and only if it admits an ordering with no backward arcs.
Several parameters exploiting an ordering with specific properties have been studied in this sense~\cite{GurskiR19} such as the cutwidth. In this problem, for each prefix of an ordering of the vertices, we associate a cut defined as the set of backward arcs with head in the prefix and tail
outside of it. This parameter then corresponds to the minimum value among all the orderings of the maximum size of any possible cut w.r.t this ordering (a formal definition is introduced in next section). It is well-known that computing cutwidth is NP-complete~\cite{gavril1977some}, and has a $O(\log^2(n))$-approximation on general graphs \cite{leighton1999multicommodity}.
Specifically on tournaments, one can compute an optimal ordering for the cutwidth by sorting the degrees according to the in-degrees~\cite{Fradkin2011}. 
%Note that we will use the same idea to show a 3-approximation algorithm to compute degreewidth\todo{not defined at this point} in \cref{th:approx}.\todo{Transition?}
%\todo{no link between sparse and degreewidth so it seems a bit odd}

In this paper, we propose a new parameter called degreewidth using the concept of backward arcs in an ordering. This parameter corresponds to the minimum value among all the orderings of the maximum number of backward arcs incident to a same vertex. Hence, an acyclic tournament is a tournament with degreewidth zero. 
Furthermore, one can notice that tournaments with degreewidth one correspond to \textit{sparse tournaments} introduced in~\cite{BessyBT17,Thiebaut19}. 
A tournament is \textit{sparse} if there exists an ordering such that the backward arcs form a matching. In particular, it is claimed in~\cite{BessyBT17} that there exists a polynomial-time algorithm for finding such ordering, but the only available algorithm appearing in~\cite[Lemma~35.1, p.97]{Thiebaut19} seems to be incomplete, as we show in this article. In~\cite{BessyBKSSTZ19} it has been proved that there is a polynomial running-time algorithm to compute a maximum sized arc-disjoint packing of triangles and also to compute a maximum sized arc-disjoint packing of cycles on sparse tournaments. On the other hand, it has been proved~\cite{BessyBT17} that packing vertex-disjoint triangles in tournaments is unlikely to admit a PTAS, even when restricted on sparse tournaments.

To the best of our knowledge this paper is the first to study this parameter.
As we will see in the next part, although having similarities with the cutwidth, this new parameter differs in certain aspects. We first study the parameter itself on both structural and computational aspects. Then, we show how this parameter can be used to solve efficiently some classical problems on tournaments.

%Finally, the notion of \textit{sparse tournaments} has been introduced in \cite{BessyBT17,Thiebaut19}. A tournament is sparse if there exists an ordering such that the backward arcs are disjoint or equivalently if it admits a fas that is a matching. In particular Bessy et al. claimed that there exists a polynomial-time algorithm for finding such ordering, but the only available algorithm appearing in~\cite[Lemma~35.1, p.97]{Thiebaut19} seems to be incomplete, as we show in this article. In~\cite{BessyBKSSTZ19} they proved that there is a polynomial running-time algorithm to compute a maximum sized arc-disjoint packing of triangles and also to compute a maximum sized arc-disjoint packing of cycles. On the other hand, it has been proved~\cite{BessyBT17} that packing vertex-disjoint triangles in tournaments is unlikely to admit a PTAS, even when restricted on sparse tournaments.

%\myparagraph{Our approach}\todo{Seems a bit redundant with results, but I kinda like this paragraph}

\myparagraph{Results and organization of the paper}

%\newJ{
	Next section gives some definitions and preliminary observations.
	In \Cref{sec:degreewidth_section}, we first study the degreewidth of a special class of tournaments, called regular tournaments, of order $2k+1$ and prove they have degreewidth $k$. We then prove that it is NP-hard to compute the degreewidth of a tournament in the general case. Fortunately, we finally give a tight 3-approximation algorithm to compute this parameter.

	Then in \cref{sec:sparse_section}, we present some results on sparse tournaments.
	We first focus on the special class of tournaments that we call $U$-tournaments. We prove there are only two possible sparse orderings for such tournaments. Then, we give a polynomial time algorithm to decide if a tournament is sparse. %\par

	Finally, in \Cref{sec:degreewidth_as_parameter} we study degreewidth as a parameter for some classical graph problems. First, we show an FPT algorithm for \textsc{Dominating Set} w.r.t degreewidth. Then, we focus on tournaments with degreewidth one. We design an algorithm running in time $O(n^3)$ to compute a \textsc{Feedback Arc Set} on sparse tournaments on $n$ vertices. However, we show that \textsc{Feedback Vertex Set} still remains NP-complete on this class of tournaments.
	Due to paucity of space the missing proofs are in Appendix.
%}

%%% Local Variables:
%%% mode: latex
%%% TeX-master: "main"
%%% End:

\section{Preliminaries}
\subsection{Notations}
\label{sec:preliminaries}
In the following, all the digraphs are simple, that is without self-loop and multiple arcs sharing the same head and tail, and all  cycles are directed cycles. The \textit{underlying graph} of a digraph $D$ is a undirected graph obtained by replacing every arc of $D$ by an edge. Furthermore, we use $[n]$ to denote the set $\{1,2, \dots, n\}$.

 Let $T$ be a tournament with vertex set $\{v_1,\dots, v_n\}$.
We denote $N^{+}(v)$ the \textit{out-neighbourhood} of a vertex $v$, that is the set $\{u \mid (v,u) \in A(T)\}$. Then, $T$ being a tournament, the \textit{in-neighbourhood} of the vertex $v$ denoted $N^{-}(v)$ corresponds to $V(T)\setminus(N^{+}(v) \cup \{v\})$.
The \textit{out-degree} (resp. \textit{in-degree}) of $v$ denoted $d^{+}(v)$ (resp. $d^{-}(v)$) is the size of its out-neighbourhood (resp. in-neighbourhood).

A tournament $T$ of order $2k+1$ is \textit{regular} if for any vertex $v$, we have $d^{+}(v) = d^{-}(v) = k$.
Let $X$ be a subset of $V(T)$. We denote by $T - X$ the subtournament induced by the vertices $V(T) \setminus X$. Furthermore, when $X$ contains only one vertex $\{v\}$ we simply write $T-v$ instead of $T-\{v\}$.
We also denote by $T[X]$ the tournament induced by the vertices of $X$.  Finally, we say that $T[X]$ \textit{dominates} $T$ if, for every $x \in X$ and every $y \in V(T)\setminus X$, we have $(x,y) \in A(T)$.
For more definitions on directed graphs, please refer to \cite{BJG08}.

%Given a tournament $T$ of order $n$, an ordering $\sigma$ of $T$ is a bijection $[n] \to V(T) $. Given a ordering $\sigma$, we can then define a strict total order $\infor$ of $V(T)$ such that for all $i$ and $j$ in $[n]$, we have $\sigma(i) \infor \sigma(j)$ if and only if $i < j$. We call the value of $\sigma^{-1}(u)$ the \textit{position} of the vertex $u$ (starting at one). Alternatively, we can see the ordering $\sigma$ as the ordered list written $\langle \sigma(1), \dots, \sigma(n) \rangle$.\smallskip

Given a tournament $T$, we equip the vertices of $T$ with is a strict total order $\infor$. This operation also defines an ordering of the set of vertices denoted by $\sigma := \langle v_1, \dots, v_n \rangle$ such that $v_i \infor v_j$ if and only if $i<j$. 
Given two distinct vertices $u$ and $v$, if $u \infor v$  we say that $u$ is \textit{before} $v$ in $\sigma$; otherwise, $u$ is \textit{after} $v$ in $\sigma$. 
Additionally, an arc $(u,v)$ is said to be \textit{forward} (resp. \textit{backward}) if $u\infor v$ (resp. $v\infor u$).
A topological ordering is an ordering without any backward arcs.
A tournament that admits a topological ordering does not contain a cycle. Hence, it is said to be \textit{acyclic}. 

A \textit{pattern} $p_1 := \langle v_1, \dots, v_k \rangle$ is a sequence of consecutive vertices in an ordering. Furthermore, considering a second pattern $p_2 := \langle u_1, \dots, u_{k'} \rangle$ where $\{v_1, \dots, v_k\}$ and $\{u_1, \dots, u_{k'}\}$  are disjoint, the pattern $\langle p_1, p_2 \rangle$ is defined by $\langle v_1, \dots, v_k, u_1, \dots, u_{k'} \rangle$.

%For an ordering $\sigma$ of $T$ and $v\in V(T)$, we denote  $\de(v)$ the number of backward arcs incident to $v$ in $\sigma$, that is $\de(v):= |\{u \mid u\infor v,\, u\in N^{+}(v)\} \cup \{u \mid v\infor u,\, u\in N^{-}(v)\}|$.

\myparagraph{\textbf{\large Degreewidth}}
Given a tournament $T$, an ordering $\sigma$ of its vertices $V(T)$ and a vertex $v\in V(T)$, we denote $\de(v)$ to be the number of backward arcs incident to $v$ in $\sigma$, that is $\de(v):= |\{u \mid u\infor v,\, u\in N^{+}(v)\} \cup \{u \mid v\infor u,\, u\in N^{-}(v)\}|$.
Then, we define the degreewidth of a tournament with respect of the ordering $\sigma$, denoted by $\dw(T):=\max\{\de(v) \mid v \in V(T)\}$. Note that $\dw(T)$ also corresponds to the maximum degree of the underlying graph induced by the backward arcs of $\sigma$.
Finally, we define the degreewidth $\DW(T)$ of the tournament $T$ as follows.%the minimum value taken by $\dw(T)$ over all the possible orderings $\sigma$. Formally, we have

\begin{definition}
	% The degreewidth of a tournament $T$, denoted $\DW(T)$, is defined as\\
	% ${\DW(T) := \min_{\sigma} \max_{v \in V(T)}\de(v)}$, where $\sigma$ is an ordering of the vertices of $T$.\\
        The degreewidth of a tournament $T$, denoted $\DW(T)$, is defined as \\
	${\DW(T) := \min_{\sigma \in \Sigma(T)} \dw(T)}$, where $\Sigma(T)$ is the set of possible orderings for $V(T)$.
\end{definition}

As mentioned before, this new parameter tries to measure how far is a tournament from being acyclic. Indeed, it is easy to see that a tournament $T$ is acyclic if and only if $\DW(T) = 0$. Additionally, when degreewidth of a tournament is one, it coincides with the notion of sparse tournaments, introduced in~\cite{BessyBT17}.

%\begin{toappendix}
%
%	\begin{observationrep}\label{obs:dw_of_subtournament}
%		Let $T$ be a tournament and $X$ a subset of its vertices.
%		Then, we have $\DW(T[X]) \leq \DW(T)$.
%	\end{observationrep}
%
%	\begin{proof}
%		Consider an optimal ordering $\sigma$ of $T$.
%		Then the restriction of this ordering to the vertices $X$, gives an ordering of $X$ such that $\dw(T[X]) \leq \DW(T)$.
%		Thus, we have $\DW(T[X]) \leq \dw(T[X]) \leq \DW(T)$.
%	\end{proof}
%\end{toappendix}

%As a direct corollary, we have if $T$ is sparse, then $T-X$ is also sparse.\todo{S:need this line? \\J:can we put it in the statement?}
%
%\todo{this following lemma is useless but I found it funny\\ J:do we remove it?}
%
%
%The following lemma will be used in the polynomial algorithm to compute a sparse ordering.

\subsection{Links to other parameters}

\myparagraph{Feedback arc/vertex set} A \textit{feedback arc set} (fas) is a collection of arcs that, when removed from the digraph (or, equivalently, reversed) makes it acyclic. Therefore, it is common to consider the size of the minimal fas as a measure if the digraph is far from being acyclic. In this context, degreewidth comes as a promising alternative. Finding a small subset of arcs hitting all substructures (in this case, directed cycles) of a digraph is one of the fundamental problems in graph theory.
\par
Note that we can easily bound the degreewidth of a tournament by its minimum fas $f$:

\begin{observation}
	For any tournament $T$, we have $\DW(T) \leq |f|$.
\end{observation}
\begin{proof}
	Consider a tournament $T$, and let $\sigma_f$ be an ordering of $T$ for which the backward arcs are exactly the $k$ arcs of a minimal feedback arc set of $T$. Then, for any vertex $v \in V(T)$, we have $\de[\sigma_f](v) \leq k$. Therefore, $\DW(T) \leq \dw[\sigma_f](T) \leq k$.
\end{proof}

\begin{figure}
	\begin{minipage}{.5\textwidth}
		\centering
		\begin{subfigure}{6.5cm}
			\centering\includegraphics[page = 2, width=0.85\linewidth]{./graphics/link_cutwidth}
			\caption{\label{fig:link_fvs_fas}Example of a tournament with degreewidth one but fas (resp. fvs) $\frac{|V(T)|}{3}$.}
		\end{subfigure}
		\begin{subfigure}{6.5cm}
			\centering\includegraphics[page = 4, width=0.85\linewidth]{./graphics/link_cutwidth}
			\vspace{-1em}
			\caption{\label{fig:link_fvs}Example of a tournament $T$ with fvs one ($v_7$) but degreewidth $\frac{|V(T)| - 3}{2}$. The topological ordering of $T - v_7$ is $\langle v_1, v_2, v_3, v_4, v_5, v_6\rangle$.}
		\end{subfigure}
	\end{minipage}%
	\begin{minipage}{.5\textwidth}
		\centering
		\begin{subfigure}{6.5cm}
			\centering\includegraphics[page = 1, width=0.85\linewidth]{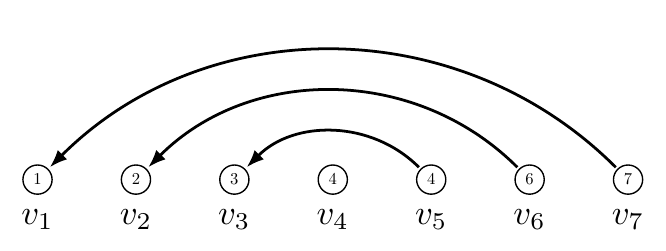}
			\caption{\label{fig:link_cutwidth}Example of a tournament with degreewidth one but cutwidth $\frac{|V(T)| - 1}{2}$. Since the vertices are sorted by increasing in-degrees (values inside the vertices), this is an optimal ordering for the cutwidth.}
		\end{subfigure}
	\end{minipage}%
	\caption{Link between degreewidth and other parameters. All the non-depicted arcs are forward.}
	\label{fig:links_with_parameters}
\end{figure}

Note however that the opposite is not true; it is possible to construct tournaments with small degreewidth but large fas, see \cref{fig:link_fvs_fas}.\smallskip

Similarly, a \textit{feedback vertex set} (fvs) consists in a collection of vertices that, when removed from the digraph makes it acyclic. However, \--- unlike the feedback arc set \--- the link between feedback vertex set and degreewidth seem less clear; we can easily construct tournaments with low degreewidth and large fvs (see  \cref{fig:link_fvs_fas}) as well as large degreewidth and small fvs (see \cref{fig:link_fvs}). \smallskip

\myparagraph{Cutwidth} Let us first recall the definition of the cutwidth of a digraph. Given an ordering $\sigma:= \langle v_1, \dots, v_n \rangle$ of the vertices of a digraph $D$, we say that a prefix of $\sigma$ is a sequence of consecutive vertices $\langle v_1, \dots, v_k \rangle$ for some $k \in [n]$. We associate for each prefix of $\sigma$ a \textit{cut} defined as the set of backward arcs with head in the prefix and tail outside of it. The \textit{width} of the ordering $\sigma$ is defined as the maximum sized
cut among all the possible prefixes of $\sigma$. Finally, the cutwidth of $D$, denoted $ctw(D)$, is the minimum width among all the orderings of the vertex set of $D$.

Intuitively, the difference between the cutwidth and the degreewidth is that the former focuses on the backward arcs going ``above'' the intervals between the vertices while the latter focuses on the backward arcs coming from and to the vertices themselves.
Observe that for any tournament $T$, the degreewidth is bounded by a function of the cutwidth. Formally, we have the following

\begin{observation}
	For any tournament $T$, we have $\DW(T) \leq 2ctw(T)$.
\end{observation}

\begin{proof}
		Consider a tournament $T$, and let $\sigma_c$ be an optimal ordering of $T$ for the cutwidth. Then, let $v$ be a vertex such that $\de[\sigma_c](v) = \dw[\sigma_c](T)$, the number of backward arcs with $v$ as a tail (resp. with $v$ as a head) cannot be larger than $ctw(T)$ without contradicting the optimality of $\sigma_c$. Therefore, we have $\dw(T) \leq \dw[\sigma_c](T)  = \de[\sigma_c](v) \leq 2 ctw(T)$.
	\end{proof}

Note however that the opposite is not true; it is possible to construct tournaments with small degreewidth but large cutwidth, see \cref{fig:link_cutwidth}.\smallskip

%\begin{lemma}\label{lem:subtourn_sparse}
%	Let $T$ be tournament. Then for any subset of vertices $X \subseteq V(T)$, if $T$ is sparse, then $T[X]$ sparse.
%\end{lemma}
%\begin{proof}
%	Let $\sigma$ be a sparse ordering of $T$, then removing from $\sigma$ the vertices of $V(T)\setminus X$ cannot create backward arcs, proving there is a sparse ordering of $T[X]$. 
%\end{proof}

%%% Local Variables:
%%% mode: latex
%%% TeX-master: "main"
%%% End:

\section{Degreewidth}
\label{sec:degreewidth_section}

In this section, we present some structural and algorithmical results for the computation of degreewidth. We first introduce the following lemma that provides a lower bound on the degreewith.

\begin{lemmarep}
	\label{lemma:degreewidth_min_indegree}
	Let $T$ be a tournament.
	Then $\DW(T) \geq \min_{v \in V(T)} d^-(v) $.
\end{lemmarep}
\begin{toappendix}
	\begin{proof}
		Consider an optimal ordering $\sigma$ of $T$.
		Denote by $u$ the first vertex according to this order. Clearly, $u$ has $d^-(u)$ backwards arcs.
		Therefore, we have $\DW(T) \geq d^-(u) \geq \min_{v \in T(V)} d^-(v)$.
	\end{proof}
\end{toappendix}

%\input{regular_tournaments}

% In this section, we prove that computing the degreewidth of a tournament is NP-complete. But first, we need to prove the following result on the degreewidth of regular tournaments.

\subsection{Degreewidth of regular tournaments}
\label{sec:regular_tournaments}

\begin{theoremrep}
	\label{th:regular_tournament}
	Let $T$ be a regular tournament of order $2k+1$. Then $\DW(T) = k$. Furthermore, for any ordering $\sigma$, by denoting $u$ and $v$ respectively the first and last vertices in $\sigma$, we have $\de(u)=\de(v)=k$.
      \end{theoremrep}

\begin{proof}
	Due to Lemma~\ref{lemma:degreewidth_min_indegree}, $\DW(T) \geq k$.
	Suppose by contradiction that $\DW(T) > k$.
	Let $\sigma$ be an ordering of $T$ such that $\dw(T) = \DW(T)$ which minimizes the total number of backward arcs.
	Let the leftmost vertex of $\sigma$ with $\de(v) > k$ be denoted by $v$.
	We construct an ordering $\sigma'$ from $\sigma$  by placing $v$ at the first position (and not moving the other vertices). First we show that  $\dw[\sigma'](T) \leq \dw(T)$.
	Since $v$ is first in $\sigma'$ and $T$ is regular, we have that $\de[\sigma'](v) = k$.
	Observe that since $T$ is regular and $\de(v)>k$, $v$ is not the first vertex in $\sigma$. Suppose that the vertex $w$ precedes $v$ in $\sigma$. Then, since $v$ is the leftmost vertex such that $\de(v)>k$, we have $\de(w) \leq k$.
	If $(v,w) \in A(T)$, then $\de[\sigma'](w) = \de(w) -1 < k$.
	Otherwise, $(w,v) \in A(T)$, then  $\de[\sigma'](w) = \de(w) + 1 \leq k+1 \leq \de(v)$.
	Since the ordering between other vertices is the same in both $\sigma$ and $\sigma'$, we have that $\dw[\sigma'](T) \leq \dw(T)$.

	Now we show that the number of backward arcs in $\sigma'$ is less than the number of backward arcs in $\sigma$ which contradicts the minimality of $\sigma$.
	Let $L^+ = N^+(v) \cap \{ u \mid u \infor v\}$ be the set of out-neighbours of $v$ on the left of $v$, $L^- = N^-(v) \cap \{ u \mid u \infor v\}$ the set of in-neighbours of $v$ on the left of $v$, $R^+ = N^+(v) \setminus L^+$ be the set of out-neighbours of $v$ on the right of $v$ and $R^- = N^-(v) \setminus L^-$ be the set of in-neighbours of $v$ on the right of $v$ in $\sigma$.
	Then $\de(v) = |L^+| + |R^-|$.
	The backward arcs from $v$ to $L^+$ are forward arcs in $\sigma'$ and the arcs from $L^-$ to $v$ are now backward arcs incident to $v$ in $\sigma'$.
	All the other arcs remain unchanged.
	As $T$ is regular, we have $|L^-| + |R^-| = k$ and then $\de(v) = |L^+| + (k-|L^-|) > k$.
	Thus, $|L^+| > |L^-|$.
	Therefore, the total number of backward arcs of $\sigma'$ is strictly smaller than $\sigma$.

	This contradicts the minimality of $\sigma$.
	Hence, we conclude that $\DW(T) = k$. The second part of the statement is immediate by regularity of the tournament.
\end{proof}

\begin{sketch}
		\Cref{lemma:degreewidth_min_indegree} guarantees us that the degreewidth of the regular tournament $T$ is at least $k$. 
		Then, we consider an ordering $\sigma$ of $V(T)$ such that $\dw(T) = \DW(T)$ which minimizes the total number of backward arcs, and we suppose that $\DW(T) > k$. 
		We obtain a contradiction by providing a new ordering  $\sigma'$ of same degreewidth but with strictly less backward arcs by moving the leftmost vertex $v$ of $\sigma$ with $d_\sigma(v) >k $ to the first position on the left. 
\end{sketch}

Note that regular tournaments contain many cycles; therefore it is not surprising that their degreewidth is large. This corroborates the idea that this parameter measures how far a tournament is from being acyclic.

\subsection{Computational complexity}
\label{sec:hardness_proof}

\noindent We now show that computing the degreewidth of a tournament by defining a reduction from \BTSAT, proven NP-complete~\cite{Berman03} where each clause contains exactly three unique literals and each variable occurs two times positively and two times negatively.

%\prob{\BTSAT}
%{a CNF formula $\varphi$ where each clause contains exactly three unique literals and each variable occurs two times positively and two times negatively.}
%{is there an assignment $\beta$ that satisfies $\varphi$? }
%\vspace{-0.1cm}
Let $\varphi$ be a \BTSAT formula with $m$ clauses $C_1,\dots,C_{m}$ and $n$ variables $x_1,\dots,x_{n}$.
Since the regular tournaments in our gadgets require an odd number of vertices, we have to ensure that $\varphi$ contains an even number of clauses and an odd number of variables. Note that if this is not the case, we can duplicate the formula $\varphi$, and add a variable that does not occur in any clause.

\begin{construction}
	\label{const:3sat}
	Let $\varphi$ be a \BTSAT formula with $m$ clauses $C_1,\dots,C_{m}$ clauses and $n$ variables $x_1,\dots,x_{n}$ such that $n$ is odd and $m$ is even. Let $W$ be an odd integer greater than $n^3 + m^3$ and such that $\frac{W+1}{2}+m+n$ is odd. We construct a tournament $T$ as follows.
	\begin{compactitem}
		\item Create two regular tournaments $A$ and $D$ of order $\frac{W+1}{2}+m+n$ such that $D$ dominates~$A$.
		\item Create two regular tournaments $B$ and $C$ of order $W$ such that $A$ dominates $B\cup C$, $B$ dominates $C$ and $B \cup C$ dominates $D$.
		\item Create an acyclic tournament $X$ of order $2n$  with topological ordering $ \langle v_1,v'_1,\dots,v_{n},v'_{n}\rangle $ such that $A\cup C$ dominates $X$ and $X$ dominates $B \cup D$.
		\item Create an acyclic tournament $Y$ of order $2m$ with topological ordering $\langle q_1,q'_1,\dots,q_{m},q'_{m}\rangle  $ such that $B \cup D$ dominates $Y$ and $Y$ dominates $A \cup C$.
		\item For each clause $C_{\ell}$ and each variable $x_i$ of $\varphi$,
		\begin{compactitem}
			\item if $x_i$ occurs positively in $C_{\ell}$, then $\{v_i,v'_i\}$ dominates $\{q_\ell,q'_\ell\}$,
			\item if $x_i$ occurs negatively in $C_\ell$, then $\{q_\ell,q'_\ell\}$ dominates $\{v_i,v'_i\}$,
			\item if $x_i$ does not occur in $C_\ell$, then introduce the paths $(v_i,q_\ell,v'_i)$ and $(v'_i,q'_\ell,v_i)$.
		\end{compactitem}
		\item Introduce an acyclic tournament $U = \{u^p_i, \bar{u}^p_i \mid i\leq n, p \leq 2\}$ of order $4n$ such that $U$ dominates $A\cup Y \cup C$ and $B\cup D$ dominates $U$. For each variable $x_i$, add the following~paths,
		\begin{compactitem}
			\item for all variable $x_k\neq x_i$ and all $p\leq 2$, introduce the paths $(v_k,u^p_i,v'_k)$ and $(v'_k,\bar{u}^p_i,v_k)$,
			\item introduce the paths $(v_i,u^1_i,v'_i)$, $(v'_i,u^2_i,v_i)$, $(v_i,\bar{u}^1_i,v'_i)$ and $(v'_i,\bar{u}^2_i,v_i)$.
		\end{compactitem}
		\item Finally, introduce an acyclic tournament $H=\{h_1,h_2\}$ with topological ordering $\langle h_1,h_2\rangle$ and such that $B \cup C \cup X \cup Y \cup D$ dominates $H$ and $H$ dominates $U$.
	\end{compactitem}
\end{construction}

We call a vertex of $X$ a \textit{variable vertex} and a vertex of $Y$ a \textit{clause vertex}. Furthermore, we say that the vertices $(v_i, v'_i)$ (resp. $(q_{\ell},q'_{\ell})$) is a \textit{pair of variable vertices} (resp. \textit{pair of clause vertices}).
\begin{definition}
	Let $T$ be a tournament resulting from \Cref{const:3sat}. An ordering $\sigma$ of $T$ is \textit{nice} if:
	\begin{compactitem}
		\item  %\todo{J: Not useful anymore, right?}
		$\dw(A)=\frac{|A|-1}{2}$, $\dw(B)=\frac{|B|-1}{2}$, $\dw(C)=\frac{|C|-1}{2}$, and $\dw(D)=\frac{|D|-1}{2}$,
		\item $\sigma$ respects the topological ordering of $U \cup Y$,
		\item $A  \infor B \infor U \infor Y \infor C \infor D \infor H$, and
		\item for any variable $x_i$, either $A \infor v_i \infor v'_i \infor B$ or $C \infor v_i \infor v'_i \infor D$.
	\end{compactitem}
\end{definition}

\begin{figure}
	\centering
	\includegraphics[width=0.90\linewidth]{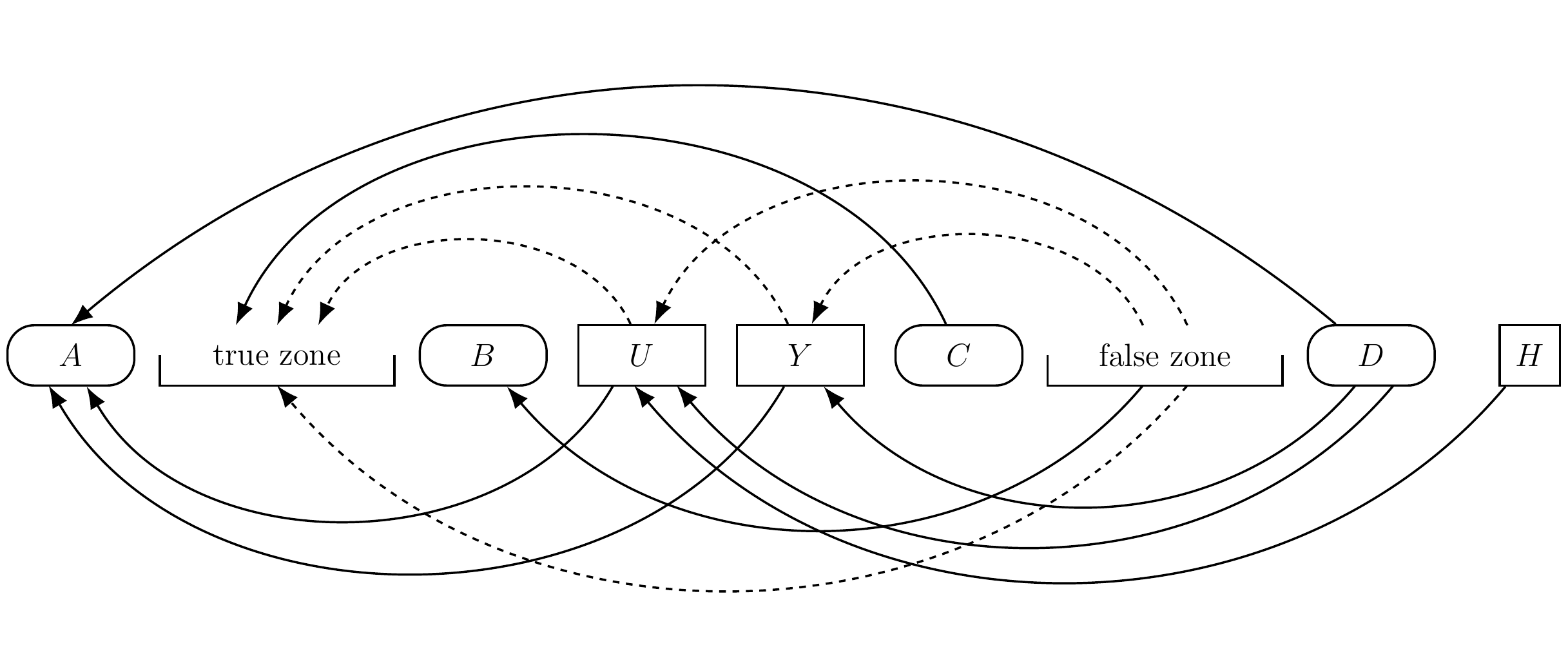}
	\caption{\label{fig:nice}Example of a nice ordering. A rectangle represent an acyclic tournament, while a rectangle with rounded corners represents a regular tournament. A plain arc between two patterns $P$ and $P'$ represents the fact that there is a backward arc between every pair of vertices $v \in P$ and $v' \in P'$. A dashed arc means some backward arcs may exist between the patterns.
		%In this example, we depict one clause vertex $q_1$ and four pairs of variables vertices $\{(v_1,v'_1),\dots,(v_4,v'_4)\}$. In the boolean formula, we have $c_1 = x_1 \vee \neg x_2 \vee x_4$. We can observe that in the depicted ordering, the pairs $(v_1,v'_1)$ and $(v_4,v'_4)$ satisfy $q_1$ whereas the pair $(v_2,v'_2)$ does not.
	}
\end{figure}

\begin{figure}
	\centering
	\includegraphics[width=0.4\textheight, page=4]{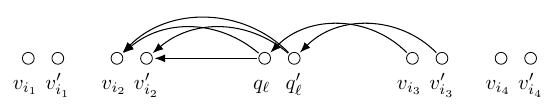}
	\caption{Example of the backward arcs for the clause $C_\ell:=x_{i_1}\vee\bar{x}_{i_2}\vee\bar{x}_{i_4}$, and $x_{i_3}$ not in $C_\ell$. We ommited the forward arcs and the potential backward arcs between two vertices of $X$.}
	\label{fig:hardness_connection}

\end{figure}

An example of a nice ordering is depicted in \Cref{fig:nice}. Let $\sigma$ be a nice ordering, we call the pattern corresponding to the vertices between $A$ and $B$, the \textit{true zone} and the pattern after the vertices of $C$ the \textit{false zone}. Let $(q_{\ell},q'_{\ell})$ be a pair of clause vertices and let $(v_i,v'_i)$ be a pair of variable vertices such that $x_i$ occurs positively (resp. negatively) in $c_\ell$ in $\varphi$. We say that the pair $(v_i,v'_i)$ satisfies $(q_{\ell},q'_{\ell})$ if $v_i$ and $v'_i$ belong both to the true zone (resp. false zone). Note that there is no backward arc between $\{q_{\ell},q'_{\ell}\}$ and $\{v_i,v'_i\}$ if and only if $(v_i,v'_i)$ satisfies $(q_{\ell},q'_{\ell})$. Notice also that for any pair of variable vertices $(v_i,v'_i)$ such that $x_i$ does not appear in $c_\ell$ and $(v_i,v'_i)$ is either in the true zone or in the false zone, then there is exactly two backwards arc between $\{q_{\ell},q'_{\ell}\}$ and $\{v_j,v'_j\}$, see \Cref{fig:hardness_connection}. \medskip

\begin{lemmarep}
	\label{lemme:nice}
	Let $T$ be a tournament resulting from \Cref{const:3sat} and let $\sigma$ be a nice ordering of $T$, we have $\dw(T) \leq W + 2m + 3n + 4$. Moreover, for any vertex $w \in V(T)\setminus Y$, we have $\de(w)<W + 2m + 3n + 4$.
\end{lemmarep}

\begin{proof}
	% A
	Let $a$ be a vertex of $A$, there are at most $\frac{|A|-1}{2}= \frac{W+1}{4}+\frac{m+n-1}{2}$ backward arcs between $a$ and $A \setminus \{a\}$. By construction, there are $|U \cup Y \cup D|=\frac{W +1}{2} +3m+5n$ backward arcs between $a$ and $V(T)\setminus A$. Thus, we have $\de(a)\leq \frac{3W+1}{4} + \frac{7m+11n}{2}< W + 2m+3n+4$. \par
	%B
	Let $b$ be a vertex of $B$, there are at most $\frac{|B|-1}{2}=\frac{W-1}{2}$ backward arcs between $b$ and $B \setminus \{b\}$. By construction, there are at most $|X|=2n$ backward arcs between $b$ and $V(T)\setminus B$. Thus, we have $\de(b)\leq \frac{W-1}{2} +2n < W + 2m+3n+4$. \par
	%C
	Let $c$ be a vertex of $C$, there are at most $\frac{|C|-1}{2}=\frac{W-1}{2}$ backward arcs between $c$ and $C \setminus \{c\}$. By construction, there are at most $|X|=2n$ backward arcs between $c$ and $V(T)\setminus C$. Thus, we have $\de(c)\leq \frac{W-1}{2} +2n < W + 2m+3n+4$.\par
	%D
	Let $d$ be a vertex of $D$, there are at most $\frac{|D|-1}{2}= \frac{W-1}{4}+\frac{m+n}{2}$ backward arcs between $d$ and $D \setminus \{d\}$. By construction, there are $|A \cup U \cup Y|=\frac{W+1}{2} +3m+5n$ backward arcs between $d$ and $V(T)\setminus D$. Thus, the degreewidth of $d$ with respects to $\sigma$ is at most $\frac{3W+1}{4} + \frac{7m+11n}{2}< W + 2m+3n+4$. \par
	% X
	Let $v$ be a vertex of $X$ such that $v \in \{v_i,v'_i\}$ for some variable $x_i$ of $\varphi$. There are at most $|X| - 1=2n - 1$ backward arcs between $v$ and $X \setminus \{v\}$. If $v$ belongs to the true zone, then there are $|C|=W$ backward arcs between $v$ and $C$ and none between $v$ and $B$. If $v$ belongs to the false zone, then there are $|B|=W$ backward arcs between $v$ and $B$ and none between $v$ and $C$. By construction there are $\frac{|U|}{2}=2n$ backward arcs between $v$ and $U$. Let $Y_i = \{q_\ell,q'_\ell \mid x_i \in c_\ell\}$, if $x_i$ occurs in the clause $c_\ell$, then there is $2$ backward arcs between $v$ and $\{q_\ell,q'_\ell\}$ and none otherwise. Since $x_i$ occurs exactly two times positively and two times negatively, there are $\frac{|Y_i|}{2}$ backward arcs between $v$ and $Y_i$. Moreover, by construction, there are exactly $\frac{|Y\setminus Y_i|}{2}$ between $v$ and $Y \setminus Y_i$. The number of backward arcs between $v$ and $Y$ is $\frac{|Y|}{2}+2=m$. Thus, we have $\de(v)\leq W +m + 4n$. Since $n<m$, $\de(v)< W +2m + 3n+4$. \par
	%U
	Let $u$ be a vertex of $U$. First, note that $\sigma$ respects the topological ordering of $U$, we have $d_{U}(u) = 0$. There are $|H|=2$ backward arcs between $u$ and $H$. There are $|A| = \frac{W+1}{2}  + m+n$ backward arcs between $u$ and $|A|$ and $|D|=\frac{W+1}{2}+m+n$ backward arcs between $u$ and $D$. Let $(v_i,v'_i)$ be a pair of variable vertices, since $v_i \infor v'_i \infor u$ or $u \infor v_i \infor v'_i$, by construction there is exactly one backward arc between $u$ and $\{v_i,v'_i\}$. Thus, there are $\frac{|X|}{2}=n$ backward arcs between $u$ and $|X|$. Hence, $\de(u) = W +2m+3n+3$. \par
	%H
	Let $h$ be a vertex of $H$. There are $|U|=4n$ backward arcs between $h$ and $U$ and none between $h$ and $V(T)\setminus U$. Thus, we have $\de(h)=4n < W + 2m+3n+4$. \par
	%Y
	Finally, let $q_\ell$ be a vertex of $Y$. By construction, there are $|A|=\frac{W+1}{2} + m +n$ backward arcs between $q_\ell$ and $A$ and $|D|=\frac{W+1}{2}+m+n$ backward arcs between $q_\ell$. Let $v_i$ and $v'_i$ be a pair of variable vertices in $X$. If $x_i$ occurs in $c_\ell$, then there are at most two backward arcs between $q_\ell$ and $\{v_i,v'_i\}$. If $x_i$ does not occur in $c_\ell$, then there is one backward arc between $q_\ell$ and $\{v_i,v'_i\}$. Thus, there are at most $n+3$ backward arcs between $q_\ell$ and $X$. Hence, we have $\de(q_\ell)\leq W + 2m + 3n +4$.
\end{proof}
In order to show the correctness of our reduction, we need to manipulate nice orderings. Hence, we introduce the following lemma to transform any ordering into a nice ordering.
\begin{lemmarep}
	\label{lemma:always nice}
	Let $T$ be a tournament resulting from \Cref{const:3sat} and let $\sigma$ be an ordering of $T$. There is a nice ordering $\sigma'$ of $T$ such that $\dw[\sigma'](T)\leq \dw(T)$.
\end{lemmarep}

\begin{proof}
	Let $\sigma$ be an ordering of $T$. First, if $\dw(T)\geq W + 2m+3n+3$, then by \Cref{lemme:nice}, for any nice ordering $\sigma'$ of $T$ we have $\dw[\sigma'](T)\leq\dw(T)$. Thus, we can suppose that $\dw(T)< W + 2m + 3n +3$. Second, if for any regular sub-tournament $T'$ among $A,B,C$ or $D$, we have $\dw(T')> \frac{|T'|-1}{2}$, then by \cref{th:regular_tournament}, we can rearrange the vertices of this tournament so that $\dw(T') \leq \frac{|T'|-1}{2}$. % Hence, we can suppose that $\sigma$ respects the property of \Cref{lemma:rotational_tournament}.
	Further, we show that it is possible to construct an ordering $\sigma'$ with $\dw[\sigma'](T)\leq \dw(T)$ and having the following properties:
	\begin{description}
		\item[Proof that $A\infor D$:]
			Let $a\in A$ be the rightmost vertex of $A$ and $d\in D$ be the leftmost vertex of $D$. Toward a contradiction, suppose that $d \infor a$. % If there is no vertex $w\in U \cup Y$ between $d$ and $a$, then by construction, we can exchange the positions of $a$ and $d$ without increasing the degreewidth of $\sigma$.
			% Suppose that there is a vertex $w \in U \cup Y$ such that $d \infor w \infor a$. If there is a vertex $t\in B \cup C$ such that $w \infor t \infor a$ (resp. $d \infor t \infor w$), then we can put $a$ just before $t$ (resp. $d$ just after $t$) in $\sigma$.
			% Hence, there no vertex of $B \cup C$ between $d$ and $a$.
			Let $BC_L = \{t \mid t \in B \cup C, t \infor a\}$ and $BC_R= \{t \mid t \in B \cup C, d \infor t\}$. Note that $BC_L \cup BC_R = B \cup C$. % (\textit{i.e.} vertices of $B \cup C$ are partitioned according to whether they are positionned to the left or right of $w$ in $\sigma$).
			If $|BC_L| > |BC_R|$, then $|BC_L|> \frac{|B \cup C|}{2} = W$. Since $A$ is a regular tournament, there are $\frac{|A|-1}{2}$ backward arcs between $a$ and $A\setminus \{a\}$. Since $BC_L \infor a$, we have $|BC_L|$ backward arcs between $BC_L$ and $a$. Hence, we have
			\begin{eqnarray*}
				\de(a) & \geq & \frac{|A|-1}{2} + |BC_L|\\
				\de(a) & \geq & \frac{W-1}{4}+\frac{m+n}{2} + W \\
				\de(a) & \geq & \frac{5W-1}{4}+\frac{m+n}{2}\\
				\de(a) & > & W + 2m+3n+4.
			\end{eqnarray*}

			We can prove similarly that if $|BC_L| \leq |BC_R|$, we also reach a contradiction. Therefore, we have $A \infor D$.

		\item[Proof that $A \infor B$ and $C \infor D$:]
			Let $a\in A$ be the rightmost vertex of $A$ and $b\in B$ be the leftmost vertex of $B$. Toward a contradiction, suppose that $b\infor a$. If there is no vertex $w \in U \cup Y$ between $b$ and $a$, then by construction, we can exchange the positions of $a$ and $b$ without increasing the degreewidth of $\sigma$. Suppose there is a vertex $w \in U \cup Y$ such that $b \infor w \infor a$. Let $B_L = \{b' \mid b' \in B, b' \infor w\}$ and $B_R=B \setminus B_L$.
			Since $A$ is a regular tournament, we have $\frac{|A|-1}{2}$ backward arcs between $a$ and $A\setminus\{a\}$. Since $B_L \infor a$, we have $|B_L|$ backward arcs between $B_L$ and $a$. By the previous item, we have $a \infor D$ and thus, there are $D$ backward arcs between $a$ and $D$. Hence, $d(a)\geq \frac{|A|-1}{2} + |B_L| + |D|$ which implies
			\begin{eqnarray*}
				\frac{|A|-1}{2} + |B_L| + |D| & < & W + 2m + 3n + 4\\
				\frac{W-1}{4} + \frac{m+n}{2} + W - |B_R| + \frac{W+1}{2} + m + n & < & W + 2m + 3n + 4\\
				|B_R| & \geq & \frac{3W}{4} -\frac{m}{2} - \frac{3n}{2} - \frac{15}{4}.
			\end{eqnarray*}
			Now consider the vertex $w$. We have $|B_R|$ backward arcs between $w$ and $B_R$. Since $w \infor a$, we have $w \infor D$ and thus, there are $|D|$ backward arcs between $w$ and $D$. We have
			\begin{eqnarray*}
				\de(w) & \geq & |B_R| + |D| \\
				\de(w) & \geq &\frac{3W}{4} -\frac{m}{2} - \frac{3n}{2} - \frac{5}{2} + \frac{W+1}{2} + m + n\\
				\de(w) & \geq &\frac{5W}{4} +\frac{m}{2} - \frac{n}{2} - \frac{13}{2} >  W + 2m + 3n +4.\\
			\end{eqnarray*}
			Since we reach a contradiction, we have $A \infor B$. By symmetry, we can use the same argument to show that $C \infor D$.

		\item[Proof that $B \infor C$:]
			Let $b \in B$ be the rightmost vertex of $B$ and $c\in C$ be the leftmost vertex of $C$. Toward a contradiction, suppose that $c \infor b$. If there is no variable vertex between $c$ and $b$, then we can exchange the positions of $c$ and $b$ without increasing the degreewidth of $\sigma$. Suppose that there is a variable vertex $v \in X$ such that $c\infor v\infor b$. Let $B_L = \{b' \mid b \in B, b \infor v\}, C_L=\{c' \mid c' \in C, c' \infor v\},B_R=B \setminus B_L$ and $C_R = C \setminus C_L$.

			Suppose there is a vertex $w \in U \cup Y$ such that $w \infor v$. Since $A \infor B$, we have  $w \infor B$ or $A \infor w$. If $w \infor B$, then by construction, we have $\de(w) \geq |B| +|D| > W + 2m +3n +3$ which is a contradiction.  If $A \infor w$, then $\de(w) \geq |A| + |B_R| + |D|$ which implies
			\begin{eqnarray*}
				|A| + |B_R| +|D| & < & W + 2m + 3n + 4\\
				W + 2m + 2n + 1 + |B_R| & < & W + 2m + 3n + 4 \\
				|B_R| & < & n +3.
			\end{eqnarray*}
			Moreover, by construction, $\de(v) \geq |B_L| + |C_R|$. Thus,
			\begin{eqnarray*}
				|B_L|+|C_R| &\leq& W + 2m+3n+ 4\\
				2W - |B_R|-|C_L| &\leq& W +  2m + 3n + 4\\
				|B_R|+|C_L| &\geq& W -2m-3n-4 \\
				|C_L| & \geq & W -2m-4n-7.
			\end{eqnarray*}
			Now, since $B$ is a regular tournament, there are $\frac{|B|-1}{2}$ backward arcs between $B\setminus \{b\}$ and $b$. By construction, we have $|C_L|$ backward arcs between $C_L$ and $b$. So,
			\begin{eqnarray*}
				\de(b) & \geq & \frac{|B|-1}{2} + |C_L| \\
				\de(b) & \geq & \frac{W-1}{2} + W -2m-4n- 7 > W +2m + 2n +4.
			\end{eqnarray*}
			By symmetry, we can use the argument to find a contradiction if there is a vertex $w\in U \cup Y$ such that $v \infor w$.

		\item[Proof that $B \infor U \infor Y\infor C$:]
			Toward a contradiction, suppose that there are two vertices $w \in U \cup Y$ and $c \in C$ such that $c \infor w$. Suppose first that $C \infor w$ then we have $\de(w)\geq |C| + |D| > W + 2m +3n +4$ which is a contradiction. Then we can partition $C$ into $C_L = \{c \mid c \in C \wedge c \infor w\}$ and $C_R = C \setminus C_L$. We know that $C_R$ is not empty and since $C \infor D$, we have $w \infor D$. Then, we have $\de(w) \geq |A| + |C_L| + |D|$ which implies
			\begin{eqnarray*}
				|A| + |C_L| + |D| & \leq & W + 2m + 3n +4\\
				2W  + 2m+2n +1 - |C_R| & \leq & W +  2m + 3n +4\\
				|C_R| &\geq & W -n -3.
			\end{eqnarray*}
			Now, as we did in the other cases, if there is no vertex $v$ between $c$ and $w$ in $\sigma$ such that $(c,v)$ and $(v,w)$ are forward arcs, we can exchange the positions of $c$ and $w$ in $\sigma$ without increasing the degreewidth. Note that here we have several cases to consider: either $v \in X$ or $w\in U$ and $v \in H$. If $v \in X$, then using the previous inequality, we obtain $\de(v) \geq |B| + |C_R| > W + 2m + 3n +4$ which is a contradiction. Now, if $w\in U$ and $v \in H$ then $\de(v) \geq  |C_R| + |D| > W + 2m + 3n +4$, also a contradiction. Therefore, we have $U \cup Y  \infor C$. \par

			By symmetry, we can show that $ B \infor U \cup Y $, using the same arguments (note however that the case where $w\in U$ and $v \in H$ does not appear). Finally, since by construction $U \cup Y$ is an acyclic tournament, we can order the vertices of $U \cup Y$ so that $U \infor Y$.

		\item[Proof that $A\infor X \infor D$:]
			If there are two vertices $a \in A$ and $v \in X$ such that $v \infor a$, then by previous items, there is no clause vertex between $v$ and $a$ in $\sigma$. Thus, we can swap the positions of $a$ and $v$ in $\sigma$ without increasing the degreewidth. That is, we can obtain an order $\sigma'$ with $A \infor X$. We prove similarly that $X \infor D$.

		\item[Proof that $D\infor H$:]
			Let $h$ be a vertex of $H$. If there is a vertex $u \in U$ such that $h \infor u$, then by the previous item $h \infor C \infor D$ and thus, $\de(h) \geq |C| + |D| > W+2m+3n+4$ which is a contradiction. Hence, we have $U < h$. By construction, we can put $h$ after $D$ in $\sigma$ without increasing the degreewidth. That is, we can obtain an order $\sigma'$ with $D \infor H$.

		\item[Proof that for any pair of variable vertices $(v_i, v'_i)$, either $ v_i \infor  v'_i \infor  B$ or $C \infor v_i\infor v'_i$:]

			First, we show that for any vertex $v \in X$, we have either $v \infor B$ or $C \infor v$ (\textit{i.e.} $v$ is either in the true zone or the false zone). We can not have $B \infor v \infor C$ since otherwise we would have $\de(v) \geq |B|+|C|\geq 2W>W+2m+3n+4$. If it exists a vertex $b \in B$ such that $b \infor v \infor C$, then any vertex between $b$ and $v$ in $\sigma$ is a variable vertex or a vertex of $B$ and, we can exchange the positions of $b$ and $v$ without increasing the degreewidth. If it exists a vertex $c \in C$ such that $B \infor v \infor c$, then any vertex between $c$ and $v$ in $\sigma$ is a variable vertex or a vertex of $C$ and, we can exchange the positions of $b$ and $v$ without increasing the degreewidth. \par
			Now, let us show that for every pair $v_i$ and $v'_i$ of variable vertices, we have $v_i \infor v'_i \infor B$ or $C \infor v_i \infor v'_i$. Note that if $v'_i\infor v_i \infor B$ or $C \infor v'_i \infor v_i$, then we can exchange the positions of $v_i$ and $v'_i$ without increasing the degreewidth.
			Let $v_i$ and $v'_i$ be a pair of variable vertices, we say that $(v_i,v'_i)$ is \emph{split} if $v_i\infor B$ and $C \infor v'_i$ or if $v'_i\infor B$ and $C \infor v_i$ (\textit{i.e.} if $v_i$ and $v'_i$ are not in the same zone).  Recall that the number of backward arcs between any vertex $u\in U$ and $V\setminus X$ is $|A|+|D|+|H|=W+2m+2n+3$. Toward a contradiction let $(v_i,v'_i)$ be a split pair. Suppose that $v_i \infor v'_i$. Then, by construction, there are exactly two backward arcs between $u^p_{i}$ and $\{v_i,v'_i\}$ and two backward arcs between $\bar{u}^p_i$ and $\{v_i,v'_i\}$. Always by construction, for each pair of variable vertices $(v_j,v'_j)$, there is exactly two backward arcs between $\{u^2_i,\bar{u}^2_i\}$ and $\{v_j,v'_j\}$ (either both $u^2_i$ and $\bar{u}^2_i$ are incident to a backward arc if $(v_j,v'_j$) is not split or one of the two vertices is incident to two backward arcs). Suppose without loss of generality that the number of backward arcs between $u^2_i$ and $X \setminus \{v_i,v'_i\}$ is bigger or equal to the number of backward arcs between $\bar{u}^2_i$ and $X \setminus \{v_i,v'_i\}$. That is, the number of backward arcs between $u^2_i$ and $X \setminus \{v_i,v'_i\}$ is at least $n-1$ and thus, the number of backward arcs between $u^2_i$ and $X$ is at least $n+1$. Hence, $\de(u^2_i)> W+2m+3n+4$ which is a contradiction. By symmetry, if $v'_i \infor v_i$, we can show that either $\de(u^1_i)> W+2m+3n+4$ or $\de(\bar{u}^1_i)> W+2m+3n+4$ which is a contradiction. Hence, no pair of variable vertices is split, that is, for each pair of variable vertices $v_i$ and $v'_i$, we have $v_i\infor v'_i \infor B$ or $C \infor v_i\infor v'_i$.
	\end{description}

\end{proof}

\noindent %Now we can show the main result of this section. 
%\todo[inline]{S: I added these lines because proof of Thorem 5 isn't here. Suggestion: add the proof in main text}
Let $\varphi$ be an instance of \BTSAT and $T$ its tournament resulting from \Cref{const:3sat}. We show that $\varphi$ is satisfiable if and only if it exists an ordering $\sigma$ of $T$ such that $\dw(T)<W + 2m +3n+4$ which yields the following theorem.

\begin{theoremrep}
	Given a tournament $T$ and an integer $k$, it is NP-complete to compute an ordering $\sigma$ of $T$ such that $\dw(T)\leq k$.
\end{theoremrep}

\begin{proof}
	Let $\varphi$ be an instance of \BTSAT and $T$ its tournament resulting from \Cref{const:3sat}. We show that $\varphi$ is satisfiable if and only if it exists an ordering $\sigma$ of $T$ such that $\dw(T)<W + 2m +3n+4$.

	First, let $\beta$ be a satisfying assignment for $\varphi$. We construct a nice ordering $\sigma$ of $T$ as follows. For each variable $x_i$, if $\beta(x_i)=\texttt{true}$ then put $v_i$ and $v'_i$ in the true zone. Otherwise, put $v_i$ and $v'_i$ in the false zone. By \Cref{lemme:nice}, for any vertex $w\not\in Y$, we have $\de(w)<W+ 2m+3n +3$. Further, let $q_\ell$ be a clause vertex. The number of backward arcs between $q_\ell$ and $V(T)\setminus X$ is equal to $|A|+|D|$. Moreover, for every variable $x_i$ that does not occur in $c_\ell$, there is exactly one backward arc between $\{v_i,v'_i\}$ and $q_\ell$. For every variable $x_i \in c_\ell$, if the value of $x_i$ in $\beta$ satisfies $c_\ell$, then there is no backward arc between $\{v_i,v'_i\}$ and $q_\ell$, otherwise there are two backward arcs between $\{v_i,v'_i\}$ and $q_\ell$. Thus, since there is at least one variable in $c_\ell$ that satisfies $c_\ell$, we have $\de(q_\ell)\leq W+ 2m+3n +2$. Hence, $\dw(T)<W + 2m+3n +4$.

	Now, let $\sigma$ be an ordering of $T$ such that $\dw(T)<W+ 2m+3n +4$. By \Cref{lemma:always nice}, we can suppose that $\sigma$ is nice. We construct an assignment $\beta$ for $\varphi$ as follows. For each variable $x_i$, if $v_i$ and $v'_i$ are in the true zone, then we set $x_i$ to true. Otherwise, if $v_i$ and $v'_i$ are in the false zone, then we set $x_i$ to false. Let $c_\ell$ be a variable of $\varphi$. Since $\de(q_\ell)<\dw(T)<W + 2m+3n +4$, there is at least one pair of variable vertices $v_i$ and $v'_i$ such that there is no backward arcs between $\{v_i,v'_i\}$ and $q_\ell$. Thus, by construction, $x_i$ satisfies $c_\ell$. Hence, $\beta$ is a satisfying assignment for $\varphi$.
\end{proof}
%\newJ[J:TODO]{
%	\begin{proofsketch}
%		
%	\end{proofsketch}
%}

%%% Local Variables:
%%% mode: latex
%%% TeX-master: "main"
%%% End:

\subsection{An approximation algorithm to compute degreewidth}
In this subsection, we prove that sorting the vertices by increasing in-degree is a tight $3$-approximation algorithm to compute the degreewidth of a tournament. Intuitively, the reasons why it returns a solution not too far from the optimal are twofold. Firstly, one can observe that the only optimal ordering for acyclic tournaments (\textit{i.e.} with degreewidth $0$) corresponds to an ordering with increasing in-degrees. Secondly, this strategy also gives an optimal solution for cutwidth in tournaments.
\begin{theoremrep}
	\label{th:approx}
	Ordering the vertices by increasing order of in-degree is a tight $3$-approximation algorithm to compute the degreewidth of a tournament.
\end{theoremrep}

\begin{figure}

	\definecolor{accent}{HTML}{fcc712}

	\centering
	\begin{subfigure}{\textwidth}
		\centering
		\begin{tikzpicture}
			\foreach[count=\x from 1] \i/\n/\y in {1/a/-1,2/b/1,2/c/-1,3/d/1,3/e/-1}{

					\pgfmathtruncatemacro\xx{(-1)^\x}
					\node[smallvertex, minimum width=0.5em] (\x) at (\x,0) [scale=0.6] {\i};
					\node at (\x,-0.45) {$v_\x$};
					%	\node[scale=0.65] at (\x,-1+\xx*0.2) {$d^+(v_\x)=\i$};
				}
			\node[smallvertex,fill=accent, minimum width=0.5em] at (4,0) [scale=0.6] {2};

			\node[scale=1] at (0,0) {$\sigma_{app}$:};

			\draw [-{Latex[length=1.8mm]}, line width = 0.8pt] (4) to [bend left = 65] (2);
			\draw [-{Latex[length=1.8mm]}, line width = 0.8pt] (5) to [bend right = 55] (4);
			\draw [-{Latex[length=1.8mm]}, line width = 0.8pt] (4) to [bend right = 55] (3);
			\draw [-{Latex[length=1.8mm]}, line width = 0.8pt] (3) to [bend right = 55] (1);

			\node[scale=1] at (7,0) {$\sigma_{opt}$:};

			\foreach[count=\x from 1] \i/\n in {1/v_1,2/v_4,2/v_2,2/v_3,3/v_5}{
					\node[smallvertex, minimum width=0.5em] (b\x) at (\x+7,0)  [scale=0.6] {\i};
					\node at (\x+7,-0.45) {$\n$};
				}

			\node[smallvertex,fill=accent, minimum width=0.5em] at (8,0) [scale=0.6] {1};
			\node[smallvertex,fill=accent, minimum width=0.5em] at (9,0) [scale=0.6] {2};
			\node[smallvertex,fill=accent, minimum width=0.5em] at (11,0) [scale=0.6] {2};
			\node[smallvertex,fill=accent, minimum width=0.5em] at (12,0) [scale=0.6] {3};

			\draw [-{Latex[length=1.8mm]}, line width = 0.8pt] (b4) to [bend right = 55] (b1);
			\draw [-{Latex[length=1.8mm]}, line width = 0.8pt] (b5) to [bend left = 55] (b2);

		\end{tikzpicture}
		\caption{Example of a tournament where the approximate algorithm can return an ordering $\sigma_{app}$ (on the left) with degreewidth three while the optimal solution is one in $\sigma_{opt}$ (on the right).}
		\label{fig:tight}
	\end{subfigure}

	\begin{subfigure}{\textwidth}
		\begin{minipage}{.5\textwidth}
			\centering
			\begin{subfigure}{7.0cm}
				\centering
				\includegraphics[width=\linewidth]{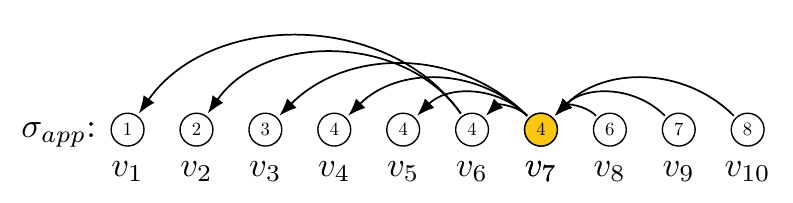}
			\end{subfigure}
		\end{minipage}%
		\begin{minipage}{.5\textwidth}
			\centering
			\begin{subfigure}{7.0cm}
				\centering
				\includegraphics[page = 2, width=\linewidth]{./graphics/approx_tight}
			\end{subfigure}
		\end{minipage}%
		\caption{Example where the optimal solution $\sigma_{opt}$ (on the right) does not have increasing in-degrees. In that case, the ratio is only $\frac{7}{3}$ with the ordering $\sigma_{app}$ the approximation algorithm can return (on the left). }
		\label{fig:not_sorted_example}
	\end{subfigure}
	\caption{Examples where the approximation solution is not optimal. The value inside a vertex corresponds to its in-degree. Coloured vertices are the ones with maximal degreewidth. Non-depicted arcs are forward.}
\end{figure}

\begin{proof}
	Let $T$ be a tournament, and consider $\sigma_{app}$ an ordering obtained by sorting the vertices of $T$ in increasing order of in-degree.  Let $v$ be a vertex such that $\de[\sigma_{app}](v)=\dw[\sigma_{app}](T)$. Similarly, denote by $\sigma_{opt}$ an optimal ordering for $T$. First, notice that if there is a vertex $u \in V(T)$ such that $3\de[\sigma_{opt}](u) \geq \de[\sigma_{app}](v)$, then $\sigma_{app}$ is a 3-approximate solution. So we can assume that for each $u\in V(T)$, we have $\de[\sigma_{opt}](u) < \frac{\de[\sigma_{app}](v)}{3}$. We consider three cases and show contradiction to this inequality in each of them.\par
	Let us first define the following sets: $D^+ = \{u \in V(T) \mid (v,u)\in A(T), u \infor[\sigma_{app}] v\}$ and $D^- = \{u \in V(T) \mid (u,v)\in A(T), v \infor[\sigma_{app}] u\}$.  Note that $\de[\sigma_{app}](v) = |D^+|+|D^-|$. Similarly, let  $R= \{u \in V(T) \mid v \infor[\sigma_{app}] u\}$ and  $L= \{u \in V(T) \mid u \infor[\sigma_{app}] v\}$. We have $d^+(v)= |D^+|+|R|-|D^-|$ and $d^-(v)= |D^-|+|L|-|D^+|$.

	Now, suppose first that $L \infor[\sigma_{opt}] v$ (\textit{i.e.} every vertex on the left of $v$ in $\sigma_{app}$ remains on the left of $v$ in $\sigma_{opt}$). We have $\de[\sigma_{opt}](v)\geq |D^+|$ which implies $2|D^+| < |D^-|$. Let $\ell$ be the leftmost vertex of $R$ in $\sigma_{opt}$. Since $d^+(\ell)\leq d^+(v)$, we have $|N^+(\ell) \cap R| \leq |D^+|+|R|-|D^-|$. Hence,
	\begin{eqnarray*}
		\de[\sigma_{opt}](\ell) &\geq& |N^-(\ell) \cap R|\\
		\de[\sigma_{opt}](\ell) &\geq& |R| - |N^+(\ell) \cap R|\\
		\de[\sigma_{opt}](\ell) &\geq& |R| - |D^+| - |R| + |D^-|\\
		\de[\sigma_{opt}](\ell) &\geq& |D^-| - |D^+|\\
		\de[\sigma_{opt}](\ell) &\geq& \frac{\de[\sigma_{app}](v)}{3}, \text{ a contradiction.}\\
	\end{eqnarray*}
	Similarly, if $v \infor[\sigma_{opt}] R$ (\textit{i.e.} every vertex after $v$ in $\sigma_{app}$ remains after of $v$ in $\sigma_{opt}$), then we can show by symmetry that for the rightmost vertex $r$ of $L$ in $\sigma_{opt}$, we have $\de[\sigma_{opt}](r) \geq  \frac{\de[\sigma_{app}](v)}{3}$, a contradiction.

	Now suppose that there is at least one vertex of $L$ after $v$ in $\sigma_{opt}$ and at least one vertex of $R$ at the left of $v$ in $\sigma_{opt}$.
	Let $M_R = \{ u \mid u \in D^+, v \infor[\sigma_{opt}] u\}$ and $M_L = \{ u \mid u \in D^-, u \infor[\sigma_{opt}] v\}$.
	Since $\de[\sigma_{opt}](v) < \frac{\de[\sigma_{app}](v)}{3}$, we have $|M_L| + |M_R| > \frac{2\de[\sigma_{app}](v)}{3}$.
	As we did before, let $\ell$ be the leftmost vertex of $R$ in $\sigma_{opt}$ and $r$ be the rightmost vertex of $L$ in $\sigma_{opt}$. Since $d^+(\ell)\leq d^+(v)$ (resp.  $d^-(r)\leq d^-(v)$), we have $|N^+(\ell) \cap (R \cup M_R)|\leq |D^+|+|R|-|D^-|$ (resp. $|N^-(r) \cap (L \cup M_L)|\leq |D^-|+|L|-|D^+|$).

	Further, since $\ell \infor[\sigma_{opt}] v$ (resp. $v \infor[\sigma_{opt}] r$), we have $\ell \infor[\sigma_{opt}] M_R \cup R \setminus \{\ell\}$ (resp. $M_L \cup L \setminus \{r\} \infor[\sigma_{opt}] r$). Hence, we have
	\begin{eqnarray*}
		\de[\sigma_{opt}](\ell) + \de[\sigma_{opt}](r)& \geq & |N^-(\ell) \cap (R \cup M_R)| + |N^+(r) \cap (L \cup M_L)|\\
		\de[\sigma_{opt}](\ell) + \de[\sigma_{opt}](r)& \geq & \big(|R|+ |M_R| - |N^+(\ell) \cap (R \cup M_R)|\big)\\
		& 	   &  + \big(|L| + |M_L|  - |N^-(r) \cap (L \cup M_L)|\big)\\
		\de[\sigma_{opt}](\ell) + \de[\sigma_{opt}](r)& \geq &  \big(|R|  + |M_R| -  |D^+|-|R|+|D^-|\big) \\
		& 	   &  + \big(|L| + |M_L|  - |D^-|-|L|+|D^+|\big)\\
		\de[\sigma_{opt}](\ell) + \de[\sigma_{opt}](r)& \geq & |M_R| + |M_L|\\
		\de[\sigma_{opt}](\ell) + \de[\sigma_{opt}](r)& \geq & \frac{2\de[\sigma_{app}](v)}{3}.\\
	\end{eqnarray*}
	Therefore, we have either $\de[\sigma_{opt}](\ell) \geq \frac{\de[\sigma_{app}](v)}{3}$ or $\de[\sigma_{opt}](r) \geq \frac{\de[\sigma_{app}](v)}{3}$, a contradiction. Finally, note that the approximation factor is tight as shown by \Cref{fig:tight}.

\end{proof}

\begin{sketch}
	We proceed by contradiction.
	Suppose that there is a tournament $T$ such that the ordering of the vertices by increasing order of the in-degrees is not a $3$-approximation of the degreewidth.
	Consider the vertex $v$ whose degreewidth in the in-degree ordering is the degreewidth of $T$ with this ordering.
	Then the degreewidth of every vertex in a optimal ordering is less than a third of the degreewidth of $v$ in the in-degree ordering.
	
	Then three cases can happen:
        \begin{inparaenum}[(a)]
        \item every vertex on the left of $v$ in the in-degree ordering remains on the left of $v$ in the optimal ordering,
        \item every vertex on the right of $v$ in the in-degree ordering remains on the right of $v$ in the optimal ordering, or
        \item neither of the two previous cases.
        \end{inparaenum}
	% Either every vertex on the left (resp. right) of $v$ in the in-degree ordering remains on the left (resp. right) of $v$ in the optimal ordering.
	% Or both previous cases do not happen.
	In these three cases, we consider vertices whose set of backward arcs can be decomposed to get a contradiction.
\end{sketch}

One could notice in \Cref{fig:tight} that the optimal ordering is also sorted by increasing in-degrees. Unfortunately, it is possible to construct examples where the optimal ordering cannot have sorted in-degrees, see \Cref{fig:not_sorted_example}.

%%% Local Variables:
%%% mode: latex
%%% TeX-master: "main"
%%% End:

\section{Results on sparse tournaments}
\label{sec:sparse_section}
In this section, we focus on tournaments with degreewidth one, called sparse tournaments.
The main result of this section is that unlike in the general case, it is possible to compute in polynomial time a sparse ordering of a tournament (if it exists).
We begin with an observation about sparse orderings (if such ordering exists) of a tournament.

\begin{lemmarep}\label{lem:pos_di}
	Let $T$ be a sparse tournament of order $n > 4$ and $\sigma$ be an ordering of its vertices. If $\sigma$ is a sparse ordering, then for any vertex $v$ such that $d^{-}(v) = i$, the only possible positions of $v$ in $\sigma$ are $\{i, i+1, i+2\} \cap [n]$.
\end{lemmarep}
\begin{proof}
	Let $\sigma$  be an ordering where there are at most $i-2$ vertices before $v$. Therefore, at least two vertices of $N^{-}(v)$ are after $v$ in $\sigma$, proving it is not a sparse ordering.

	Similarly, if we consider an ordering $\sigma$ where there are at least $i+3$ vertices before $v$. Therefore, at least two vertices of $N^{+}(v)$ are before $v$ in $\sigma$, proving it is not a sparse ordering.
\end{proof}

Note that \cref{lem:pos_di} gives immediately an exponential running-time algorithm to decide if a tournament is sparse. However, we give in \cref{subsec:poly_sparse} a polynomial running-time algorithm for this problem. Before that we study a useful subclass of sparse tournaments, we call the $U$-tournaments.
\subsection{$U$-tournaments}
\label{subsec:Un_tournaments}

\begin{figure}
	\begin{minipage}{.5\textwidth}
		\centering
		\begin{subfigure}{6.5cm}
			\centering\includegraphics[page = 1, width=0.85\linewidth]{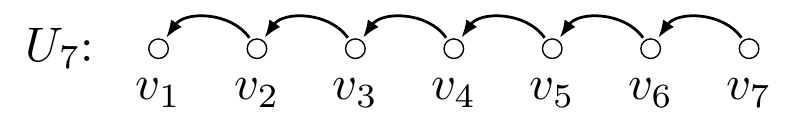}
			\caption{The tournament $U_7$.}
			\label{fig:u7}
		\end{subfigure}
		\begin{subfigure}{6.5cm}
			\centering\includegraphics[page = 3, width=0.85\linewidth]{./graphics/Un}
			\caption{The sparse ordering $\Pi_1(U_7)$. Note that $v_1$ is the only vertex not incident to any backward arc.}
			\label{fig:pi_1_u7}
		\end{subfigure}
		\begin{subfigure}{6.5cm}
			\centering\includegraphics[page = 2, width=0.85\linewidth]{./graphics/Un}
			\caption{The sparse ordering $\Pi_7(U_7)$. Note that $v_7$ is the only vertex not incident to any backward arc.}
			\label{fig:pi_7_u7}
		\end{subfigure}
	\end{minipage}%
	\begin{minipage}{.5\textwidth}
		\centering
		\begin{subfigure}{6.5cm}
			\centering\includegraphics[page = 6, width=0.85\linewidth]{./graphics/Un}
			\caption{The tournament $U_8$.}
			\label{fig:u8}
		\end{subfigure}
		\begin{subfigure}{6.5cm}
			\centering\includegraphics[page = 4, width=0.85\linewidth]{./graphics/Un}
			\caption{The sparse ordering $\Pi(U_8)$. The dashed forward arcs is a minimum feedback arc set of the tournament. Note that all the vertices are incident to one backward arc. }
			\label{fig:pi_u8}
		\end{subfigure}
		\begin{subfigure}{6.5cm}
			\centering\includegraphics[page = 5, width=0.85\linewidth]{./graphics/Un}
			\caption{The sparse ordering $\Pi_{1,8}(U_8)$. Note that $v_1$ and $v_8$ are the only vertices not incident to any backward arc.}
			\label{fig:pi_1_8_u8}
		\end{subfigure}%
	\end{minipage}
	\caption{The tournaments $U_7$ and $U_8$ and their sparse orderings. The non-depicted arcs are forward arcs.}
	\label{fig:u7_u8}
\end{figure}

In this subsection, we study one specific type of tournaments called $U$-tournaments. Informally, they correspond to the acyclic tournaments where we reversed all the arcs of its Hamiltonian path.
\begin{definition}
	For any integer $n \geq 1$, we define $U_n$ as the tournament on $n$ vertices with $V(U_n)= \{v_1,v_2, \dots, v_n\}$ and $A(U_n) = \{(v_{i+1}, v_i) \mid $ $\forall i \in [n-1]\} \cup \{(v_i, v_j) \mid 1\leq i <n, i+1 <  j \leq n\}$. We say that a tournament of order $n$ is a $U$-tournament if it is isomorphic to $U_n$.
\end{definition}
\Cref{fig:u7,fig:u8} depict respectively the tournaments $U_7$ and $U_8$. These tournaments seem somehow strongly related to sparse tournaments and the following results will be useful later for both the polynomial algorithm to decide if a tournament is sparse and the polynomial algorithm for minimum feedback arc set in sparse tournaments.
To do so, we prove that each $U$-tournament of order $n > 4$ has exactly two sparse orderings of its vertices that we formally define.

\begin{definition}
	Let $P(k) = \langle v_{k+1},v_{k} \rangle$ be a pattern of two vertices of $U_n$ for some integer $k \in [n-1]$.
	For any integer $n \geq 2$, we define the following special orderings of $U_n$:
	\begin{compactitem}
		\item if $n$ is even:
		\begin{compactitem}
			\item $\Pi(U_n)$ is the ordering given by $\langle v_1, P(2), P(4), \dots, P(n-2), v_{n}\rangle$.
			\item $\Pi_{1,n}(U_n)$ is the ordering given by $\langle P(1), P(3), \dots, P(n-2),P(n)\rangle$.
		\end{compactitem}
		\item if $n$ is odd:
		\begin{compactitem}
			\item $\Pi_{1}(U_n)$ is the ordering given by $\langle P(1), P(3), \dots, P(n-2), v_n\rangle $.
			\item $\Pi_{n}(U_n)$ is the ordering given by $ \langle v_1, P(2), P(4), \dots, P(n-3),P(n-1)\rangle $.
		\end{compactitem}
	\end{compactitem}
\end{definition}
\medskip

\Cref{fig:pi_1_u7,fig:pi_7_u7} (and \cref{fig:pi_u8,fig:pi_1_8_u8}) depict the orderings $\Pi_1(U_7)$ and $\Pi_7(U_7)$ (resp. $\Pi(U_8)$ and $\Pi_{1,8}(U_8)$) of the tournament $U_7$ (resp. $U_8$).
One can notice that these orderings are sparse and the subscript of $\Pi$ indicates the vertex (or vertices) without a backward arc incident to it in this ordering. In the following, we prove that when $n>4$ there are no other sparse orderings of $U_n$. Note, however, that there are three possible sparse orderings of $U_3$ (namely, $\Pi_1(U_3)$ and $\Pi_3(U_3)$ defined previously, as well as $\Pi_2(U_3):=\langle v_3, v_2, v_1\rangle$)  and three sparse orderings of $U_4$ (namely, $\Pi(U_4)$, $\Pi_{1,4}(U_4)$ as defined before, and $\Pi'(U_4):=\langle v_2, v_4, v_1, v_3\rangle$).

\begin{toappendix}
	In order to prove that there are no other sparse orderings of $U_n$, we start by giving some properties on the position of the vertices; specifically, we refine the statement of \Cref{lem:pos_di} in the case where the tournament is $U_n$.

	\begin{lemmarep}\label{lem:posv1vn}
		In any sparse ordering $\sigma$ of $U_n$, the position of $v_1$ (and $v_n$) is either $1$ or $2$ (resp. $n$ or $n-1$). Furthermore, there are no pattern $\langle v_i,v_{i+1}\rangle$ in $\sigma$ for each $i \in [n-1]$.
	\end{lemmarep}

	\begin{proof}
		We prove the first statement for the vertex $v_1$. Using \Cref{lem:pos_di}, we already know that $v_1$ is either at position 1, 2 or 3. Suppose the latter, so there are exactly two vertices before $v_1$. By construction, one of these two vertices has to be $v_2$ and let $v_k$ be the other vertex before $v_1$, where $k \geq 3$. If in addition we have $k < n$, then let us consider the vertex $v_{k+1}$ which is after $v_{k}$ in $\sigma$. Therefore, we have $\de(v_k) \geq 2$, proving $\sigma$ is not a sparse ordering. If $k = n > 4$, then $v_3$ is after $v_n$, so we also have $\de(v_n) \geq 2$. The proof for the vertex $v_n$ is similar. \par\medskip

		Let us now proof that there no two consecutive vertices $\langle v_i,v_{i+1}\rangle$ for each $i \in [n-1]$.
		By contradiction, consider a sparse ordering $\sigma$ such that $v_i$ and $v_{i+1}$ are consecutive. By definition of $U_n$, the arc $(v_{i+1}, v_{i})$ is a backward arc. Suppose first that $i>1$. Since $\sigma$ is sparse, then the in-neighbours of $v_{i}$ (resp. $v_{i+1}$) are exactly the vertices before $v_{i}$ (resp. $v_{i+1}$). So the vertex $v_{i-1}$ is necessarily between $v_{i}$ and $v_{i+1}$, yielding a contradiction.  \par
		Let us consider now the case $i = 1$. Note that if $v_1$ is not the first vertex, then $v_k$ for some $k\geq 3$ is before $v_1$, contradicting \Cref{lem:pos_di}. Then $v_3$ is after $v_2$, proving that $\de(v_2) \geq 2$, a contradiction.
	\end{proof}
\end{toappendix}
%\begin{lemmarep}\label{lem:2segment}
%In any sparse ordering of $U_n$ there are no two consecutive vertices $\langle v_i,v_{i+1}\rangle$ for each $i \in [n-1]$.
%\end{lemmarep}
%
%\begin{proof}
%\end{proof}

\begin{theoremrep}
	\label{thm:un_only_two_sparse_orderings}
	For each integer $n > 4$ there are exactly two sparse orderings of $U_n$. Specifically, if $n$ is even, these two sparse orderings are $\Pi(U_n)$ and $\Pi_{1,n}(U_n)$; otherwise, the two sparse orderings are $\Pi_{1}(U_n)$ and $\Pi_{n}(U_n)$.
\end{theoremrep}
\begin{proof}
	We prove the theorem by induction on the number of vertices. First, we show that $\Pi_{1}(U_5)$ and $\Pi_{5}(U_5)$ are the two only sparse tournaments of $U_5$. Using \Cref{lem:posv1vn}, we know that $v_1$ is either at position 1 or position 2 in any sparse ordering. Suppose the former. \Cref{lem:posv1vn} forbid the vertex $v_2$ to be after $v_1$, then the only possible vertex at position 2 is $v_3$ and the only possible remaining position for $v_2$ is the third one. Finally, we cannot have the pattern $\langle v_4, v_5 \rangle$ by \Cref{lem:posv1vn}, so the only possible sparse ordering of $U_5$ with $v_1$ in first position is $\langle v_1, v_3, v_2, v_5, v_4 \rangle$, that is $\Pi_{5}(U_5)$. \par
	Similarly, suppose now $v_1$ is at position 2. Then the first vertex is necessarily $v_2$. Note that $v_3$ cannot be at position 3 since it would have two backward arcs $(v_3, v_2)$ and $(v_4, v_3)$. Then the only other option by  \Cref{lem:posv1vn} is $v_4$. Then we necessarily obtain the ordering $\langle v_2, v_1, v_4, v_3, v_5 \rangle$, that is $\Pi_{1}(U_5)$.

	Similarly, we prove that $\Pi(U_6)$ and $\Pi_{1,6}(U_6)$ are the two only sparse tournaments of $U_6$. Let us suppose that $v_1$ is the first vertex. Then, as before $v_3$ is at position 2, and $v_2$ and position 3. Note that $v_4$ cannot be at position 4 since it would have two backward arcs $(v_5, v_4)$ and $(v_4, v_3)$. Then the last possible position for $v_4$ is 5, which leads to the ordering  $\langle v_1, v_3, v_2, v_5, v_4 , v_6 \rangle$, that is $\Pi(U_6)$.\par
	Finally, if we suppose that $v_1$ is at position 2, using the same arguments as for $\Pi_{1}(U_5)$ we directly obtain the ordering $\Pi_{1,6}(U_6)$.\par \medskip

	Suppose now that $U_n$ respects the statement of the theorem, and let us prove that $U_{n+1}$ does too. Let us first consider the case where $n$ is even. Note that if we remove the vertex $v_{n+1}$ from $U_{n+1}$, we obtain exactly $U_n$. Now, since $n$ is even, consider the ordering $\Pi(U_n)$ on which we will insert $v_{n+1}$. By \Cref{lem:posv1vn} we can only insert $v_{n+1}$ at position $n$, and we obtain exactly $\Pi_{n+1}(U_{n+1})$. If we now consider the ordering $\Pi_{1,n}(U_n)$ on which we will also insert $v_{n+1}$, then by \Cref{lem:posv1vn} we can only insert $v_{n+1}$ at position $n+1$, and we obtain exactly $\Pi_{1}(U_{n+1})$. This concludes the case where $n$ is even.	\par

	Let us now suppose $n$ odd. Similarly as before, note that if we remove the vertex $v_{n+1}$ from $U_{n+1}$, we obtain exactly $U_n$. Consider first the ordering $\Pi_{n}(U_n)$ on which we will insert $v_{n+1}$. By \Cref{lem:posv1vn} we can only insert $v_{n+1}$ at position $n+1$, and we obtain exactly $\Pi(U_{n+1})$. If we now consider the ordering $\Pi_{1}(U_n)$ on which we will also insert $v_{n+1}$, then by \Cref{lem:posv1vn} we can only insert $v_{n+1}$ at position $n$, and we obtain exactly $\Pi_{1,n+1}(U_{n+1})$.

	Since in every case, the vertex $v_{n+1}$ had no other possible position it proves that there are no other sparse orderings of $U_{n+1}$, concluding the proof.

\end{proof}
%\newJ[TO CHECK]{
	\begin{sketch}
		The theorem is proven by induction on the number of vertices. First, we can easily check the base cases that $U_5$ and $U_6$ both have only two sparse orderings as desired.

		Then, we assume that $U_n$ admits only two sparse orderings. Note that if we remove $v_{n+1}$ from $U_{n+1}$, we obtain $U_n$ where we can use our induction hypothesis.
		Now, let us first consider the case where $n$ is even.
		Thus, $U_n$ admits the sparse ordering $\Pi(U_n)$ on which we insert $v_{n+1}$.
		By \Cref{lem:posv1vn} we can only insert $v_{n+1}$ at position $n$, and we obtain exactly $\Pi_{n+1}(U_{n+1})$.
		Similarly, for the second ordering $\Pi_{1,n}(U_n)$, we can only insert $v_{n+1}$ at position $n+1$, leading to $\Pi_{1}(U_{n+1})$ which concludes the case $n$ even.
		We then do a similar reasoning for $n$ odd.

		%Then, we prove the induction hypothesis by considering the case where the number of vertices is odd and even. Namely, we prove in both cases there is only two ways to insert an extra vertex in the previous sparse orderings.
	\end{sketch}

\subsection{A polynomial-time algorithm for sparse tournaments}
\label{subsec:poly_sparse}

We give here a polynomial algorithm to compute a sparse ordering of a tournament (if any). First of all, let us recall a classical algorithm to compute a topological ordering of a tournament (if any): we look for the vertex $v$ with the smallest in-degree; if $v$ has in-degree one or more, we have a certificate that the tournament is not acyclic. Otherwise, we add $v$ at the beginning of the ordering, and we repeat the reasoning on $T - v$, until $V(T)$ is empty. 

The idea of the algorithm to compute a sparse ordering is similar; we consider iteratively the vertices with low in-degrees. 
In most cases, there is a trivial recursion.
However, unlike the algorithm for the topological ordering, we may have to look more carefully how the vertices with low in-degrees are connected to the rest of the digraph. 
These correspond to the case where there exists a $U$-sub-tournament of $T$ which either dominates or ``quasi-dominates'' (defined later) the tournament $T$.
Because of the latter possibility (where a backward arc $(a,b)$ is forced to appear), we need to look for specific sparse orderings, called $M$-sparse orderings ($a$ or $b$ should not be end-vertices of other backward arcs).
As all the $U$-tournament sparse orderings have been described, we can derive a recursive algorithm.

% \begin{definition}
%  $T$ is $M$-sparse if there exists an ordering $\sigma$ of $T$ such that $\dw \leq 1$ and $\de(v) = 0$ for all $v \in M$.
% \end{definition}
% \todo[inline]{S: Let $X$ be a sub-tournament of a tournament $T$ and $M \subseteq V(T)$. We say $X$ is $M$-sparse if there exists an ordering $\sigma$ of $X$ such that $\dw(X) \leq 1$ and $\de(v) = 0$ for all $v \in M$.}
\begin{definition}
Let $T$ be a tournament, $X$ be a subset of vertices of $T$, and $M$ be a subset of $X$. We say $T[X]$ is $M$-sparse if there exists an ordering $\sigma$ of $X$ such that $\Delta_{\sigma(T[X])}(X) \leq 1$ and $\de(v) = 0$ for all $v \in M$. In that case, $\sigma$ is said to be a $M$-sparse ordering of $T[X]$. 
\end{definition}
For example, $U_4[\{v_1,v_2,v_3\}]$ is $\{v_2\}$-sparse, because there exists a sparse ordering (for instance $\sigma := \langle v_3,v_2,v_1\rangle$) of $U_4[\{v_1,v_2,v_3\}]$ such that $\de(v_2) = 0$.
We remark that $T$ is sparse if and only if $T$ is $\emptyset$-sparse. As a matter of fact, the algorithm described in this section computes a $\emptyset$-sparse ordering of the given tournament (if any).
\begin{toappendix}
\begin{observationrep}
    \label{lemma:restriction_M_sparse}
    Let $T$ be a tournament and $X$ and $M$ be subsets of vertices of $T$.
    If $T$ is $M$-sparse, then $T[X]$ is $M \cap X$-sparse.
\end{observationrep}
\begin{proof}
    Consider $\sigma$ an ordering of the vertices of $T$ such that it is $M$-sparse.
    The restriction of $\sigma$ to the vertices of $X$ is also a sparse ordering, and $\de(v) = 0$ for all $v \in M$.
    Thus $T[X]$ is also $M \cap X$-sparse.
\end{proof}

\begin{lemmarep}
    \label{lemma:case_dominating}
    Let $T$ be a tournament, let $X$ and $M$ be two subsets of $V(T)$ such that $X$ dominates $T$. Then, $T$ is $M$-sparse if and only if $T[X]$ is $M \cap X$-sparse and $T-X$ is $M \setminus X$-sparse.
\end{lemmarep}
\begin{proof}
    Suppose that $T[X]$ is $M\cap X$-sparse and that $T-X$ is $M \setminus X$-sparse.
    Then by concatenating the orders, we obtain a $M$-sparse ordering for $T$.
    This follows from the fact that we do not create any additional backwards arcs by concatenating since $X$ dominates $T-X$.
    
    Suppose that $T$ is $M$-sparse.
    Then by \cref{lemma:restriction_M_sparse}, $T[X]$ is $M \cap X$-sparse and $T-X$ is $M \setminus X$-sparse.
\end{proof}

\begin{corollary}
\label{lemma:poly_case0}
    Let $T$ be a tournament and $v$ be a vertex such that $d^-(v) = 0$.
    Let $M$ be a subset of $V(T)$.
    Then $T$ is $M$-sparse if and only if $T-v$ is $M \setminus \{v\}$-sparse.
\end{corollary}
%\begin{proof}
%    Suppose that $T$ is $M$-sparse, then $T-v$ is $M$-sparse by \cref{lemma:restriction_M_sparse}.
%    
%    \noindent Suppose now that the subtournament $T-v$ is $M$-sparse. Then there exists a $M$-sparse ordering $\sigma$ of $T-v$.
%    As $d^-(v) = 0$, then we consider the ordering $\sigma'$ where $v$ is in the first position, that is $\sigma' = \langle v, \sigma \rangle$. This is trivially a sparse ordering of $T$, which does not depends whether $v \in M$ or not.
%\end{proof}
%

\begin{lemmarep}
    \label{lemma:case_one_indegree_1}
    Let $T$ be a tournament such that there exists a unique vertex $v$ with $d^-(v) = 1$ and all the other vertices have in-degree at least two. 
    Let $w$ be the unique in-neighbour of $v$ and $M$ be a subset of vertices of $V(T)$.
    Then $T$ is $M$-sparse if and only if $v \not\in M$ and $T-v$ is $M\cup \{w\} \setminus \{v\}$-sparse.
\end{lemmarep}

\begin{proof}
    Suppose first that $T$ is $M$-sparse. Note that in any sparse ordering, the first vertex is necessarily $v$ otherwise any vertex placed at the first would have two backward arcs incident to it, that is, the ordering would not be sparse. Therefore, in any sparse ordering, there is a backward arc from $w$ to $v$.  Thus, we have $v \not\in M$. 
     Consider now a $M$-sparse ordering $\sigma := \langle v, \sigma' \rangle$ of $T$.
    Then, $\sigma'$ is also a sparse ordering of $T-v$.
    Furthermore, notice that we have $\dw[\sigma'](w) = 0$, as there is already a backwards arc from $w$ to $v$ in $\sigma$.
    Thus $\sigma'$ is a $M\cup \{w\} \setminus \{v\}$-sparse ordering of $T-v$.
    
    For the other directions, suppose that $v\not\in M$ and $T-v$ is $M\cup \{w\} \setminus \{v\}$-sparse and let $\sigma'$ be a $M\cup \{w\} \setminus \{v\}$-sparse ordering of $T-v$.
    Consider now the following ordering $\sigma := \langle v, \sigma' \rangle$ of $T$. 
    Note that $\sigma$ is sparse since there is only one backward arc incident to $v$, namely $(w,v)$.
    Therefore, $\sigma$ is a $M$-sparse ordering since $\sigma'$ is a $M\cup \{w\} \setminus \{v\}$-sparse ordering.
\end{proof}
\end{toappendix}

\begin{figure}
	\centering
	\includegraphics[width=0.35\textheight, page=2]{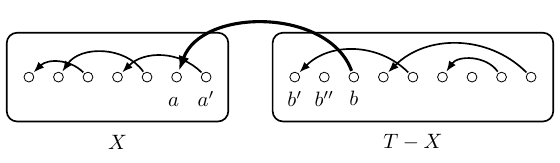}
	\caption{An example where $X$ $(b,a)$-quasi-dominates $T$. Non-depicted arcs are forward. The vertex $a'$ is an out-neighbour of $a$ in $X$, and $b'$, $b''$ are in-neighbours of $b$ in $T-X$.}
	\label{fig:quasi_dom}
\end{figure}

\begin{definition}[see \Cref{fig:quasi_dom}]
    Given a tournament $T$ and two of its vertices $a$ and $b$, we say that a subset of vertices $X$ \textit{quasi-dominates} $T$ if: 
    \begin{compactitem}
      \item there exists an arc $(b,a) \in A(T)$ such that $a \in X$ and $b \not\in X$,
      \item $(u,v) \in A(T)$ for every $(u,v) \in ( X \times (V(T)\setminus X)) \setminus \{(a,b) \}$,
      \item $d^-(b) \geq |X| +1$, and 
      \item the vertex $a$ has an out-neighbour in $X$.
    \end{compactitem}
    In this case, we also say $X$ $(b,a)$-\textit{quasi-dominates} $T$.
\end{definition}

% See \Cref{fig:quasi_dom} for an example of a quasi-dominating set.
\begin{toappendix}

\begin{lemmarep}
    \label{lemma:case_quasi_dominating}
Let $T$ be a tournament, $X$ be a subset of its vertices of $T$, and $a$ and $b$ be two vertices such that $X$ $(b,a)$-quasi-dominates $T$. 
    Furthermore, let $M$ be a subset of $V(T)$.     
    Then $T$ is $M$-sparse if and only if $T[X]$ is $(M \cup \{a\}) \cap X$-sparse and $T-X$ is $(M \cup \{b\}) \setminus X$-sparse
\end{lemmarep}
\begin{proof}
    Suppose first that $X$ is $(M \cup \{a\}) \cap X$-sparse and that $T-X$ is $(M \cup \{b\}) \setminus X$-sparse (see \Cref{fig:quasi_dom_M_sparse} for an example). We want to define a $M$-sparse ordering of $T$. To do so, let $\sigma'$ be a $(M \cup \{a\}) \cap X$-sparse ordering of $X$ and $\sigma''$ be a $(M \cup \{b\}) \setminus X$-sparse ordering of $T-X$.
    We define the ordering of $T$, let $\sigma:= \langle \sigma', \sigma''\rangle$.
    Note that $\sigma$ is a sparse ordering. Indeed, for every vertex $v$ different from $a$ and $b$, we have $\de(v) \leq 1$. Furthermore, we also have $\de(a) = \de(b) = 1$ since $(b,a) \in A(T)$ and there is no backward arc incident to $a$ in $\sigma'$ and there is  no backward arc incident to $b$ in $\sigma''$.\par\smallskip
    
    \begin{figure}
	\centering
	\includegraphics[width=0.35\textheight]{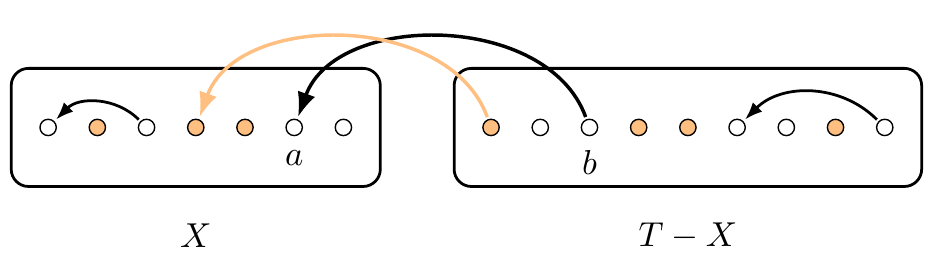}
	\caption{Example of a tournament $T$ where $X$ $(b,a)$-quasi-dominates $T$ and $X$ is $(M \cup \{a\})\cap X$-sparse and $T-X$ is $(M\cup\{b\})\setminus X$-sparse. Vertices of $M$ are coloured orange.}
	\label{fig:quasi_dom_M_sparse}
    \end{figure}
    
    Suppose now that $T$ is $M$-sparse, and consider $\sigma$ a $M$-sparse ordering of $T$.
    If $a \infor b$, then $(b,a)$ is a backward arc and $\de(a) = \de(b) = 1$.
    Therefore, the restriction of $\sigma$ to $X$ is $(M \cup \{a\}) \cap X$-sparse (as $b \not\in X$).
    Furthermore, the restriction of $\sigma$ to $T-X$ is $(M \cup \{b\}) \setminus X$-sparse (as $a \not\in V \setminus X $). So we proved the statement in this case. 
    
    Let us now consider the case $b \infor a$. 
    As $d^-(b) \geq |X|+1$ and as every vertex of $X$ except $a$ is an in-neighbour of $b$, then there exists two vertices $b'$ and $b''$ in $V\setminus X$ such that $(b',b) \in A(T)$ and $(b'',b) \in A(T)$.
    Note that since $\de(b) \leq 1$, either $b'$ or $b''$ must be before $b$ in the ordering. Without loss of generality, suppose that $b' \infor b$.
    By definition, $a$ as a out-neighbour in $X$, call it $a'$. Then $(a',b') \in A(T)$.
    
    If $a \infor a'$, then $b'$ has at least two backward arcs: $(a',b')$ and $(a,b')$, which contradicts the ordering $\sigma$ being sparse. Thus $a' \infor a$ and so $a$ has at least two backward arcs: $(a,a')$ and $(a,b)$. We also reach a contradiction, proving the case $b \infor a$ is impossible, and concluding the proof.
\end{proof}

\begin{definition}
    Let $T$ be a tournament and $X = (v_1, \ldots, v_k)$ be a list of vertices with $k \geq 2$.
    We say that $X$ satisfies the $U$-property if $d^-(v_1) = 1$ and for each $i \in \{2,\dots, k\}$, we have     $(v_i, v_{i-1}) \in A(T)$ and $d^-(v_i) = i-1$.
\end{definition}

\begin{lemmarep}
    \label{lemma:degree_to_Un}
    Let $T$ be a tournament and $X$ be a list of vertices satisfying the $U$-property.
    Then $T[X]$ is the tournament $U_k$.
\end{lemmarep}

\begin{proof}
    We will prove by induction the following assertion: the subtournament $T[\{v_1,\ldots, v_i\}]$ is $U_i$ for any $i \geq 1$.
    The assertion is true for $i = 1$.
    Suppose that it is true for $i \geq 1$.
    Let us prove that it is true for $i+1$.
    Let $1\leq j < i$.
    As $v_1, \ldots, v_i$ is $U_i$, then $v_1, \ldots, v_{j-2}, v_{j+1}$ are the in-neighbours of $v_j$ which is of in-degree $j-1$.
    Thus, $v_{i+1}$ is a out-neighbour $v_j$, that is, $(v_j,v_{i+1}) \in A(T)$ for any $1 \leq j < i$.
    We deduce that $T[\{v_1,\ldots, v_{i+1}\}]$ is $U_{i+1}$, proving the statement.
\end{proof}

%\begin{lemma}
%    Let $T$ be a tournament such that there exists $\langle u_1, \ldots, u_k\rangle$ a pattern of $ k \geq 2$ vertices such that for every $i \in \{2,\dots, k\}$, we have $(u_i, u_{i-1}) \in A(T)$, $d^-(u_i) = i-1$ and $d^-(u_1) = 1$. 
%    Then:
%    $u_k$ has another in-neighbour called $u_{k+1}$ (different from $u_1, \ldots, u_k$) with $d^-(u_{k+1}) \geq k-1$, $T[\{u_1, \ldots, u_{k+1}\}]$ is $U_{k+1}$ and
%    \begin{itemize}
%    \item if $d^-(u_{k+1}) = k-1$, then $\{u_1, \ldots, u_{k+1}\}$ is dominating $T$
%    \item if $d^-(u_{k+1}) \geq k+1$, then $\{u_1, \ldots, u_{k}\}$ is $(u_k,u_{k+1})$-quasi-dominating $T$.
%    \end{itemize}
%\end{lemma}
%
%\begin{proof}
%    Due to \cref{lemma:degree_to_Un}, the restriction of $T$ to $\{u_1, \ldots, u_{k+1}\}$ is $U_{k+1}$. \todo{can previous Lemma apply? What if $d^-(u_{k+1}) \neq k$?}
%    Furthermore, we have $d^-(u_{k+1}) \geq k-1$.
%    
%    Consider first the case where $d^-(u_{k+1}) = k-1$.
%    Let $v$ be  a vertex of $V(T) \setminus \{u_1, \ldots, u_{k+1}\}$, then $(u_i,v) \in A(T)$ for every $i$.
%    Then $\{u_1, \ldots, u_{k+1}\}$ is dominating $T$.
%
%    Suppose now that $d^{-}(u_{k+1}) \geq k+1$.
%    As $T[\{u_1, \ldots, u_{k+1}\}]$ is the tournament $U_{k+1}$, then $\{u_1, \ldots, u_{k}\}$ is $(u_k,u_{k+1})$-quasi dominating $T$ (because $u_k$ has a out-neighbour in $\{u_1, \ldots, u_k\}$ which is $u_{k-1})$ and because $d^-(u_{k+1}) \geq |\{u_1, \ldots, u_k\}|+1$).
%\end{proof}

\begin{observation}  Let $T$ be a tournament and a list $X = (v_1, \ldots, v_k)$ of vertices satisfying the $U$-property.. Then  $v_k$ has one in-neighbor in $V(T) \setminus X$. 
\end{observation}
\begin{proof}
    Since $X$ satisfies the $U$-property, we have $d^-(v_k) = k-1$ and $\{v_1, \ldots, v_{k-2}\} \subset N^-(v_k)$ because of $T[X] = U_k$. Thus, we deduce that $|N^-(v_k) \setminus X| = 1$.
\end{proof}

\begin{lemmarep}
        \label{lemma:generating_Usubtournament}
    Let $T$ be a tournament and a list $X = (v_1, \ldots, v_k)$ of vertices satisfying the $U$-property.
    Let $w$ be the vertex of $N^-(v_k) \setminus X$. We denote $(v_1, \ldots, v_k, w)$ by $X'$.
    \begin{compactitem}
    \item If $d^-(w) = d^-(v_k)$, then $X'$ is a $U$-sub-tournament dominating $T$.
    \item If $d^-(w) = d^-(v_k)+1$, then $X'$ is included in a $U$-sub-tournament dominating or quasi-dominating $T$.
    \item If $d^-(w) > d^-(v_k)+1$, then $X$ is a $U$-sub-tournament $(w,v_k)$-quasi-dominating $T$.
    \end{compactitem}
    Remark that in every cases $X$ is included in a $U$-sub-tournament dominating or quasi-dominating $T$.
\end{lemmarep}
\begin{proof}

    We prove this lemma by induction on $k$ the number of vertices of $X$.
    If $k = n-1$, then $X'$ is a $U$-sub-tournament dominating $T$.
    Suppose now that the result is true for $k+1$.
    We will prove that it is true for $k$.

    Observe that by \cref{lemma:degree_to_Un}, $T[X] = U_k$.

%    Let us show that $w$ is well defined.
%    As $d^-(v_k) = k-1$ and as $\{v_1, \ldots, v_{k-2}\} \subset N^-(v_k)$ because of $T[X] = U_k$, we deduce that $|N^-(v_k) \setminus X| = 1$.
%    Thus, $w$ is well defined.
    
    Suppose that $d^-(w) = d^-(v_k)$.
    Let $v$ be  a vertex of $V(T) \setminus (X \cup \{w\})$.
    Let $i \in [k]$. Since $T[X]$ is $U_k$, we have that $d^-(v_i) = i-1$, if $i > 1$, and $d^-(v_1) = 1$, otherwise. Then the in-neighbors of $v_i$ are in $X \cup \{w\}$. 
    Thus, $v$ is a out-neighbor of $v_i$, i.e., $(v_i,v) \in A(T)$.
    As $d^-(w) = d^-(v_k)$, the in-neighbors of $w$ are in $X \cup \{w\}$.
    Thus, $(v_i,w) \in A(T)$.
    We deduce that $\forall v \in V(T) \setminus (X \cup \{w\})$, $(u,v) \in A(T), \forall u \in X \cup \{w\}$.
    Therefore, $X \cup \{w\}$ dominates $T$.
    
    Suppose that $d^-(w) = d^-(v_k)+1$.
    Then $d^-(w) = k-1 + 1 = (k+1)-1$ and we deduce that $X' = (v_1, \ldots, v_k, w)$ satisfies the $U$-property.
    Thus, by induction, $X'$ is included in a $U$-sub-tournament dominating or quasi-dominating $T$.
    
    Suppose that $d^-(w) > d^-(v_k)+1$.
    Let us show that $X$ $(w,v_k)$-quasi dominates $T$. That is, we show that the four conditions for quasi-domination holds.
    First, the arc $(w,v_k) \in A(T)$ and $w \not\in X$.
    Let $i \in [k-1]$. Note that $d^-(v_i) = i-1$, if $i > 1$, and $d^-(v_i) = 1$, otherwise. As $T[X] = U_k$, we have that $V(T)\setminus X \subset N^+(v_i)$.
    As $d^-(v_k) = k-1$ and as $T[X] = U_k$, then $V(T)\setminus (X \cup \{w\}) \subset N^+(v_k)$. 
    Thus, $(u,v) \in A(T)$ for every $(u,v) \in \{X \times (V(T)\setminus X)\} \setminus \{(v_k,w)\}$.
    Furthermore, $d^-(w) > d^-(v_k)+1 = k-1 + 1 = |X|$ and $v_k$ has a out-neighbor in $X$ which is $v_{k-1}$.
    We deduce that $X$ $(w,v_k)$-quasi-dominates $T$.
\end{proof}

\end{toappendix}

We can create the algorithm \texttt{isUkMsparse} (in appendix) which given $(v_1, \ldots, v_k)$ a $U$-tournament and $M$ a subset of these vertices, returns a boolean which is True if and only if this tournament is $M$-sparse. 
We can also create the algorithm \texttt{getUsubtournament} (in appendix) which given $T$ a tournament, and $X = (u_1, \ldots, u_k)$  a list of vertices such that $d^-(u_1) = 1$ and $d^-(u_i) = i-1$ and $(u_i,u_{i-1}) \in A(T)$ for all $i \in \{2, \ldots, k\}$, returns a $U$-subtournament dominating or quasi-dominating $T$.

With these two previous algorithms, we can derive \Cref{algo:sparse} \texttt{isMsparse}.

\begin{toappendix}

    %\begin{minipage}[t]{0.48\linewidth}\normalem
    \begin{algorithm}[H]\normalem
    \SetKwFunction{getUsubtournament}{getUsubtournament}
    %\small
    \KwData{$T$ a tournament, and $X=(u_1, \ldots, u_k)$ a list of vertices such that $d^-(u_1) = 1$ and $d^-(u_i)=i-1$ and $(u_i, u_{i-1})\in A(T)$ for all $i \in \{2, \ldots, k\}$.}
    \KwResult{ A $U$-subtournament dominating or quasi-dominating $T$.}
    %\Begin{
        $w \longleftarrow$  a vertex of $N^-(u_k) \setminus X$\;
        \lIf{$d^-(w) = d^-(u_k)$}{
            \KwRet{$X \cup \{w\}$} \tcc*[f]{ this set dominates $T$} 
        }
        \lElseIf{$d^-(w) = d^-(u_k) + 1$}{
            \KwRet{\getUsubtournament{$T, X\cup\{w\}$}}
        }
        \lElse{
            \KwRet{$X$} \tcc*[f]{ this set $(w,u_k)$-quasi-dominates $T$} 
        }
            % }
    \caption{\label{algo:getUsubtournament}\texttt{getUsubtournament}}
    \end{algorithm}
    
    %\end{minipage}\hfill\vline\hfill

    %\begin{minipage}[t]{0.48\linewidth}\normalem
    \begin{algorithm}[H]\normalem
    \SetKwFunction{isUkMsparse}{isUkMsparse}
    \KwData{$(v_1, \ldots, v_k)$ a $U_k$ tournament, $M$ a subset of the vertices of $U_k$}
    \KwResult{\texttt{True} if $U_k$ is $M$-sparse and \texttt{False} otherwise}
    %\Begin{
        \lIf{ $k \leq 2$}{
            \KwRet{\texttt{True}}
        }
        \lElseIf{$k =3$}{
            \KwRet{$|M| \leq 1$}\
        }
        \lElseIf{$k$ is even}{
            \KwRet{ $| M \setminus \{ v_1,v_k\}| = 0 $}\
        }
        \lElseIf{$k$ is odd}{
            \KwRet{ ($v_1 \not\in M$ or $v_k \not\in M$) and  $| M \setminus \{ v_1,v_k\}| = 0 $}
        }
    %}
    \caption{\label{algo:isUkMsparse}\texttt{isUkMsparse}}
    \end{algorithm}
    %\end{minipage}

    \end{toappendix}

\begin{algorithm}[t]\normalem\scriptsize
    \SetKwFunction{isMsparse}{isMsparse}
    \KwData{$T$ a tournament, $M$ a subset of the vertices of $T$}
    \KwResult{\texttt{True} if $T$ is $M$-sparse and \texttt{False} otherwise}
    %\Begin{
        \lIf{$|V(T)| \leq 1$}{
            \KwRet{\texttt{True}}
        }
        \lElseIf{$\min_{v \in V(T)} d^-(v) \geq 2$}{
            \KwRet{\texttt{False}}
        }
        \uElseIf{$\min_{v \in V(T)} d^-(v) = 0$}{
            $v \longleftarrow$  the vertex of in-degree $0$\;
            \KwRet{\isMsparse{$T -v,M \setminus \{v\}$}}\;
        }
        \uElseIf{$|\{v \in V(T) : d^-(v) = 1\}| = 1$}{
            $v,w \longleftarrow$ two vertices such that $d^-(v)=1$ and $(w,v) \in A(T)$\;
    %        $v \longleftarrow$ the vertex of in-degree $1$\;
    %        $w \longleftarrow$ the in-neighbour of $v$\;
            \KwRet{$v \not\in M$ and \isMsparse{$T-v$, $(M \cup \{w\})\setminus \{v\}$} }\;
        }
        \Else{
            $v,w \longleftarrow$ two vertices of in-degree $1$ such that $(w,v) \in A(T)$\;
            $X \longleftarrow$ \texttt{getUsubtournament}($T$,$(v,w)$)\;
            \uIf{$X$ dominates $T$}{
                \KwRet{(\texttt{isUkMsparse}($X$,$M \cap X$) and \isMsparse{$T-X$, $M\setminus X$})}\;
            }
            \Else{
                $a,b \longleftarrow$ the vertices such that $X$ $(b,a)$-quasi-dominates $T$\;
                \KwRet{(\texttt{isUkMsparse}($X$,$(M \cup \{a\}) \cap X $) and \isMsparse{$T-X$, $(M \cup \{b\})\setminus X$})}\;
            }
        }
    %}
    \caption{\label{algo:sparse}\texttt{isMsparse}}
    \end{algorithm}

\begin{theoremrep}\label{th:polysparse}
  \Cref{algo:sparse} is correct. Hence, it is possible to decide if a tournament $T$ with $n$ vertices is sparse in $\mathcal{O}(n^3)$ by calling \texttt{isMsparse($T$,$\emptyset$)}.
\end{theoremrep}

\begin{proof}

Let us show that \Cref{algo:isUkMsparse} is correct.
If $k \leq 2$, then $U_2$ is $M$-sparse for any subset $M$ of vertices of $U_2$ as there exists an ordering of $U_2$ without backward arcs (line 1).
If $k = 3$, there are only $3$ sparse orderings of $U_3$.
As the description of these sparse orderings has been seen, we can see that $U_3$ is $M$-sparse if and only if $M$ contains at most $1$ vertex (line 2).
If $k \geq 4$ and $k$ is even, we have showed that there exists sparse orderings where every vertex is adjacent to a backward arc and exactly one other sparse ordering where only $v_1$ and $v_k$ are not adjacent to a backward arc.
Thus $U_k$ is sparse if and only if there does not exist $i \in \{2, \ldots, k-1\}$ such that $v_i \in M$.
In other words $U_k$ is $M$-sparse if and only if $M \setminus \{v_1,v_k\} = \emptyset$ (line 3).
If $k \geq 4$ and $k$ is odd, we have showed that there exists exactly two sparse orderings of $U_k$: one where $v_1$ is the only vertex not adjacent to a backward arc and one another where $v_k$ is the only vertex not adjacent to a backward arc.
Thus $U_k$ is $M$-sparse if and only if $M$ does not contain both $v_1$ and $v_k$ (otherwise none of the two previous sparse orderings fit the condition) and $M$ does not contain a vertex $v_i$ such that $i \in \{2, \ldots, k-1\}$.
In other words $U_k$ is $M$-sparse if and only if $\{v_1, v_k\} \not\subset M$ and $M \setminus \{v_1, v_k\} = \emptyset$ (line 4). Thus, we show that for each value of $k \in [n]$, \Cref{algo:isUkMsparse} correctly decides if $U_k$ is $M$-sparse.

\Cref{algo:getUsubtournament} is correct by \cref{lemma:generating_Usubtournament}.

    Let us show that \Cref{algo:sparse} is correct.
  If $T$ is constituted by a single vertex then $T$ is trivially sparse (line 1).
  If  $\min_{v \in V(T)} d^-(v) \geq 2$, then by \cref{lemma:degreewidth_min_indegree}, $T$ is not sparse (line 2).
  If $T$ has a vertex $v$ of in-degree zero, then by \cref{lemma:poly_case0}, $T$ is $M$-sparse if and only if $T-v$ is $M \setminus \{v\}$-sparse (lines 5).
  Otherwise, there exists a vertex $v$ such that $d^-(v) = 1$.
    If $v$ is the unique vertex of in-degree one, then by \cref{lemma:case_one_indegree_1}, $T$ is $M$-sparse if and only if $v\not\in M$ and $T-v$ is $(M \cup \{b\}) \setminus \{v\}$-sparse (where $b$ is the unique in-neighbour of $v$) (line 9).
    Otherwise, there exist at least two vertices $v$ and $w$ of in-degree one.
    %, and suppose without loss of generality that $(v,w)\in A(T)$. 
    By \cref{lemma:generating_Usubtournament}, there exists $X$ such that either $X$ dominates $T$, or $X$ quasi-dominates $T$.
    If $X$ dominates $T$, then $T$ is $M$-sparse if and only if $X$ is $M \cap X$-sparse and $T-X$ is $M \setminus X$-sparse due to \cref{lemma:case_dominating} (line 14).
    Otherwise, there exists two vertices $a$ and $b$ such that $X$ $(b,a)$-quasi-dominates $T$, then by \cref{lemma:case_quasi_dominating}, $T$ is $M$-sparse if and only if $X$ is $(M\cup \{a\}) \cap X$-sparse and $T-X$ is $(M\cup \{b\}) \setminus X$-sparse (line 17).
    
    \vspace{0.5mm}
    
    Let us now investigate the time complexity of the algorithms.
    
    First we show that \Cref{algo:isUkMsparse} runs in time $O(n)$.
    As $M$ has size at most $n$ and computing $|M|$ and $|M \setminus \{v_1, v_k\}|$ runs in time $O(M)$ and thus the total time is $O(n)$.
    
    Let us now show that \Cref{algo:getUsubtournament} runs in $O(n^2)$.
    As $N^-(u_k)$ is of size at most $n$, then finding $w$ (line 1) can be done in time $O(n)$.
    Computing the in-degree of $w$ costs $O(n)$.
    The in-degree of $u_k$ is $k-1$ by definition.
    According to the master theorem of analysis of algorithm, this algorithm runs in $O(n^2)$.
    
    Let us now show that \Cref{algo:sparse} runs in $O(n^3)$.
    Computing $\min_{v \in V(T)} d^-(v)$ and finding the vertices which minimizes the in-degree runs in $O(n^2)$. 
    % O(n) is possible by only investigating for low in-degree with a data structure where we have the in-neighbors lists
    The vertices $(a,b)$ in Line 15 such that $X$ $(b,a)$-quasi-dominates $T$ can be computed during \Cref{algo:getUsubtournament} and thus it results in an empty cost.
    All the other operations runs in $O(n^2)$.
    According to the master theorem of analysis of algorithm, this algorithm runs in $O(n^3)$.
\end{proof}

\noindent Observe that we can easily modify \Cref{algo:sparse} to obtain a sparse ordering (if exists). Next corollary follows from the above algorithm.

\begin{corollary}
    The vertex set of a sparse tournament on $n$ vertices can be decomposed into a sequence  $U_{n_1}, U_{n_2}, \dots, U_{n_\ell}$  for some $ \ell \leq n$ such that 
each $T[U_i]$ dominates or quasi-dominates  $T[\underset{i<j\leq \ell}{\cup} U_{n_j}]$ and $\sum_{i\in [\ell]} n_i = n$.
\end{corollary}

%%% Local Variables:
%%% mode: latex
%%% TeX-master: "main"
%%% End:

\section{Degreewidth as a parameter}
\label{sec:degreewidth_as_parameter}
\subsection{Dominating set parameterized by degreewidth}
A set of vertices $X$ of a directed graph is a \textit{dominating set (DS)} if for each vertex $v \in V(v)\setminus X$, we have $N^+(v)\cap X \neq \emptyset$. 
%\newS{ 
Observe that in graphs where degreewidth is zero, DS is of size one. Similarly, for tournaments with degreewidth equals to one, the DS is of size at most two. That is, we have trivial solutions for DS for acyclic and sparse tournaments. This motivates us to look for FPT algorithm parameterized by degreewidth.
%}
%\todo{S: should say these lines here?\\J:I'm fine with them here. Where else did you have in mind?} 
In the following, we develop a FPT algorithm for {\sc Dominating Set} using a colour coding technique. Before that we observe that size of a dominating is always bounded by the size of degreewidth.

\begin{observation}\label{obs:DSlessDW}
  The size of a minimum size dominating set of a tournament $T$ is less than $\DW(T)+1$.
\end{observation}
\begin{proof}
  Consider an ordering $\sigma$ of $T$ such that $\DW_\sigma(T)$ is the degreewidth of $T$. Then, the first vertex $v$ in $\sigma$ dominates every vertex except the ones from which there is an backward arc incident to it. Therefore,  $\{v \} \cup N^-(v)$ is a dominating set of $T$. Since $v$ is the first vertex in $\sigma$, the size of $N^-(v)$ is bounded by degreewidth. Hence, the statement follows.
\end{proof}
\begin{theoremrep}
  {\sc Dominating Set} is FPT in tournaments with respect to degreewidth.
\end{theoremrep}
\begin{proof}
  Let $T$ be a tournament with degreewidth bounded by some integer $k$. We want to compute a dominating set of $T$ of size at most $s$. Using \Cref{th:approx}, we can find a $3$-approximation for degreewidth. Let $\sigma$ be the ordering given by \Cref{th:approx}. Therefore, we have $\DW_\sigma(T) \leq 3k$.\par

  Our algorithm proceeds in two steps as described below.
  First is the separation	phase where we define a  subgraph of $T$ and use $n$-$p$-$q$-{\it lopsided universal family} to identify a solution. Next, we verify it.
  To state the algorithm formally, we first define $n$-$p$-$q$-{\it lopsided universal family}.

  Given a universe $U$ and an integer $i$, we denote all the $i$-sized subsets of $U$ by ${U \choose i}$. We say that a  family
  $\mathcal{F}$ of sets over a universe $U$ with $\vert U\vert=n$, is an $n$-$p$-$q$-{\it lopsided universal family} if for every $A \in {U \choose p}$ and $B \in {U \setminus A \choose q}$, there is an $F \in \mathcal{F}$ such that $A \subseteq F$ and $B \cap F = \emptyset$.

  \begin{lemma}[\cite{FominLPS16}]
    \label{lem:lopsidedUniversal}
    There is an algorithm that given $n,p,q\in {\mathbb N}$ constructs an  $n$-$p$-$q$-lopsided universal family $\mathcal{F}$ of cardinality ${p+q \choose p} \cdot 2^{o(p+q)}  \log n$ in time
    $\vert  \mathcal{F} \vert  n$.
  \end{lemma}

  Let $|V(T)| =n$. We fix an arbitrary ordering of the vertices $V(T)$ and write $V(T)$ as $[n]$ and for $X \subseteq [n]$, we write $T[X]$ to denote the tournament induced on $X$.
  The algorithm described as follows.
  \begin{enumerate}
    \item  For each integer $1\leq p \leq s$, we construct a $n$-$p$-$3kp$-lopsided universal family $\mathcal{F}$ using the algorithm in \Cref{lem:lopsidedUniversal}.
    \item Then, for each $F \in \mathcal{F}$,
    let  $C_1, \dots, C_\ell$ be the strongly connected components of $T[F]$, ordered according to their first vertex in $\sigma$ (\textit{i.e.} the first vertex of $C_i$ is before all the vertices of $C_j$ in $\sigma$ for each $j>i$).  Check if $C_1$ is a dominating set for $T$. If so, we return $C_1$, otherwise it is a no-instance.
  \end{enumerate}

  We now show the correctness of our algorithm. Suppose $(T,s)$ is a yes-instance. Let $S$ denote a dominating set of size $s$ of $T$. Let $N = \{v \in N^+(S) \setminus S \mid v \infor u, \textit{ for some } u \in S\}$. Observe that $|N| \leq 3ks$.
  From the definition of $n$-$s$-$3ks$-lopsided universal family, we have that
  there exists a set $F \in \mathcal{F}$ such that
  \begin{align}
    S \subseteq F, and\hfill \label{it:ds} \\
    N \cap F =\emptyset.\hfill\label{it:non-solnNgs}
  \end{align}
  Now we show that if $C_i \cap S \neq \emptyset$ for some $i \in [\ell]$, then $C_i \subseteq S$. Suppose not. Let $v \in C_i \cap S$, and let $u$ be a vertex of $C_i \setminus S$. Since $C_i$ is strongly connected, let $(u:=v_1, v_2, \dots, v_p:=v)$ be a path from $u$ to $v$ in $C_i$. The vertex $u \in F$, so it is not in $N$ by \eqref{it:non-solnNgs}. Furthermore, it is not in $S$ by definition. So by definition of $N$, $u$ is not incident to any vertex of $S$. So $v_2 \in C_i \setminus S$. By repeating this reasoning, we obtain $v \notin S$, a contradiction.
  
  Finally, we show that $C_1$ is a dominating set of $T$ of size at most $s$.
  Suppose not. Let $C_i$ be the first strongly connected component in $S$ for some $i > 1$.
  Note that given two distinct strongly connected components $C_j$ and $C_{j'}$ with $j<j'$, there is, by definition, no arc between them in $T[F]$, and therefore in $T$.
  So there is no backward arcs from $C_{j'}$ to $C_j$ in $T$.
  This observation shows that $C_i$ does not dominates the vertices of $C_1, \dots, C_{i-1}$ in $T$.
  Similarly, the vertices of $C_1, \dots, C_{i-1}$ cannot be dominated by any vertices of $C_j$ for any $j>i$. So $S$ is not a dominating set of $T$, a contradiction.
  Therefore, we can return the vertices of $C_1$ as an solution of {\sc Dominating Set} in $T$.
  The algorithm invokes \Cref{lem:lopsidedUniversal} $s$ times.
  Hence, it runs in time $2^{O(s \log(s(3k+1))}n^{O(1)}$.
  Finally, \Cref{obs:DSlessDW} gives the theorem.

\end{proof}

\begin{sketch}
	We use following key observation to obtain the algorithm. If the degreewidth is bounded, a vertex in the solution can dominate bounded number of vertices that precede it. This enables us to use ``random separation'' method to separate a solution vertex from its out-neighbours that precede it. Hence, we get a randomized FPT algorithm. We use standard tool called universal family to present a de-randomized algorithm.
\end{sketch}

%%% Local Variables:
%%% mode: latex
%%% TeX-master: "main"
%%% End:

\subsection{\FAST is fast in sparse tournaments}
\label{subsec:fast_poly}
A \textit{forbidden pattern} corresponds to the patterns $\Pi(U_{2k})$ for any $k \geq 1$ as well as $\Pi'(U_4) := \langle v_2, v_4, v_1, v_3\rangle$. An example of the forbidden pattern $\Pi(U_8)$ is depicted in \Cref{fig:pi_u8}.
We say a sparse ordering has \textit{forbidden pattern} if a forbidden pattern appear as a contiguous subsequence of the ordering.
Intuitively, the problem of such patterns is that the set of their backward arcs is not a minimum fas. Hopefully, we can use \Cref{thm:un_only_two_sparse_orderings} in such a way that if the pattern $\Pi(U_{2k})$ appears, we can restructure it into  $\Pi_{1,2k}(U_{2k})$.

\begin{lemmarep}
	\label{lemma:no_forbidden_configuration}
	Let $T$ be a sparse tournament on $n$ vertices. Then, it is possible to construct in time $O(n^3)$ a sparse ordering for $T$ without forbidden patterns.
\end{lemmarep}
\begin{proof}
	Let $\sigma$ be a sparse ordering of $T$ where for some $2 \leq 2k \leq n$, the vertices $\{v_1, \dots, v_{2k}\}$ form the forbidden pattern $\Pi(U_{2k})$ (or $\Pi'(U_4)$). That is, $\sigma := \langle \sigma_1, \Pi(U_{2k}), \sigma_2 \rangle$ (resp. $\langle \sigma_1, \Pi'(U_4), \sigma_2 \rangle$). Let $\sigma'$ be the ordering we get by replacing $\Pi(U_{2k})$  by $\Pi_{1,2k}(U_{2k})$. That is, $\sigma' := \langle \sigma_1,
		\Pi_{1,2k}(U_{2k}), \sigma_2 \rangle$.  Observe that $\sigma'$ is a sparse ordering.
	Let us now show that there is no vertex of ${v_1, \dots, v_{2k}}$ lying in a forbidden pattern in $\sigma'$.
	By contradiction, suppose that $\sigma'$ has a forbidden pattern $\Pi(U_{2k'})$ (or $\Pi'(U_4)$) for some $2 \leq 2k' \leq n$ on the subset of vertices $V'$.

	\textbf{Case 1:} $\{v_1, \dots, v_{2k}\} \subseteq V'$. This is not possible since it is easy to see that the pattern $\Pi_{1,2k}(U_{2k})$ cannot be contained in the pattern $\Pi(U_{2k'})$ (resp. $\Pi' (U_4)$).

	\textbf{Case 2:} $|\{v_1, \dots, v_{2k}\} \cap V'| \geq 1$. Then, since the patterns are consecutive sequence of vertices, either the first or the last vertex of $\Pi_{1,2k}(U_{2k})$ is in $V'$. Without loss of generality, suppose that the first vertex of $\Pi_{1,2k}(U_{2k})$ is in $V'$, that is, $v_2 \in V'$. Note that $v_2$ has a backward arc incident to it in $\Pi_{1,2k}(U_{2k})$. Since $\sigma'$ is a sparse ordering, $v_2$ cannot have another backward arc to/from a vertex in $V'$. Hence, $v_2$ can not form the forbidden pattern $\Pi(U_{2k'})$ (resp. $\Pi'(U_4)$) in $V'$.

	Hence, in both cases, we have a contradiction. Thus, no forbidden pattern containing a vertex from ${v_1, \dots, v_{2k}}$ is in $\sigma'$. Therefore, given a sparse ordering $\sigma$ of $T$, we can replace the forbidden patterns and obtain a sparse ordering of $T$ containing no forbidden patterns. Next we show that we can do it in time $O(n)$. The correctness of the following process follows from the above argument.

	Given a sparse ordering  $\sigma:= \langle v_1, v_2, \dots, v_n \rangle $, we first check there is no arc $(v_{i+1},v_{i})$. If so, we swap these vertices in the ordering. Then, we check similarly for the pattern $\Pi'(U_4)$ for every four consecutive vertices and replace then with $\Pi_{1,4}(U_4)$. We can now assume $\sigma$ is a sparse ordering without these patterns.

	Now, we define the \textit{span} of an arc in an ordering $\sigma$  to be the number of vertices between its end-vertices in $\sigma$, including the end-vertices. For example, let  $\sigma:= \langle v_1, v_2, \dots, v_n\rangle$, then in $\Pi(U_{2k})$ for some $k\geq 2$, the span of the arc $(v_3,v_1)$ is three. Note that the sequence of backward arcs in $\Pi(U_{2k})$ (taken from left to right) starts and ends with backward arcs of span of three, with (eventually) backward arcs of span four in between. The idea of the following algorithm is to look for such sequences.

	We try to look for the sequence $\Pi(U_{2k})$ from left, for some $k\geq 2$.
	We check for the backward arc $(v_{s+2},v_s)$ of span three with minimum position $s$. Then, we continue to look for backward arcs of span four and stop at a backward arc of span three as described next.
	We continue as long as there is the arc $(v_{s+2i+4},v_{s+2i+1}) \in A(T)$ for $i \geq 0 $. Suppose that we end at $i = t$ such that $s+2t+4 < n$, then we check if $(v_{s+2t+5}, v_{s+2t+3}) \in A(T)$.

	If so, we have found the forbidden pattern $\Pi(U_{2t+6})$ on the vertices $\{v_s, \dots, v_{s+2t+5}\}$. We reorder this pattern according to the order $\Pi_{1,2t+6}(U_{2t+6})$ in $\sigma$.  We repeat the process from the vertex $v_{s+2t+7}$ by checking for a backward arc of span three.

	If not, we repeat the process starting from the vertex $v_{s+2t+5}$.
	Hence, we replace all the forbidden patterns in time $O(n)$ by the above left to right scan. Since by \Cref{th:polysparse}, a sparse ordering $\sigma$ of $T$ can be constructed in $O(n^3)$ time, we have proved the lemma.
\end{proof}

Let us now prove that the set of backward arcs of a sparse ordering without forbidden patterns is a minimum fas; this implies that we can compute a fas in polynomial time in sparse tournaments.
\begin{theoremrep}
	\label{lemma:minfast_poly_sparse}
	\FAST is solvable in time $O(n^3)$ in sparse tournaments on $n$ vertices.
\end{theoremrep}

\begin{proof}
	Let $T$ be a sparse tournament and let $\sigma$ be a sparse ordering without forbidden patterns of $V(T)$ computed using \cref{lemma:no_forbidden_configuration}.
	We prove that the set of backward arcs of $T$ in the ordering $\sigma$ is a minimum feedback arc set of $T$. In the following, let $B=((u_1,v_1),\dots,(u_k,v_k))$ be the set of backward arcs defined by the ordering $\sigma$. The set $B$ is ordered from the left to right according to the head of the arcs, that is, the arc $(u_i,v_i)$ appears before the arc $(u_j,v_j)$ if $v_i\infor v_j$. Let $S$ be any feedback arc set. To show that $B$ is a minimum feedback arc set, we construct an injective function $f : B \to S$ in the following way.
	We start with the function $f : B \to \{ \emptyset \}$ and then we assign iteratively a backward arc of $B$ to an arc of $S$ according to the order of $B$ from $(u_1,v_1)$.

	Let $(u_i,v_i)$ be a backward arc of $B$ to assign (all the backward arcs $(u_j,v_j)$ with $j<i$ have already been assigned). Let $x_i$ be the vertex right after $v_i$ in $\sigma$. We have $x_i\neq u_i$ since otherwise $\langle u_i,v_i\rangle$ would be isomorphic to the forbidden pattern $\Pi(U_2)$. Thus, $(v_i,x_i,u_i)$ is a cycle (as $\sigma$ is a sparse ordering) and there is at least one arc in $S$ among $(v_i,x_i), (x_i,u_i)$, and $(u_i,v_i)$ (as $S$ is a feedback arc set). We consider the four following cases.
	\begin{enumerate}[(a)]
		\item If $(u_i,v_i)\in S$, then we set $f((u_i,v_i)):=(u_i,v_i)$\label{a}.
		\item If $(u_i,v_i) \not\in S$ and $(x_i,u_i)\in S$, then we set $f((u_i,v_i)):=(x_i,u_i)$\label{b}.
		\item If $(u_i,v_i)\not\in S$, $(x_i,u_i)\not\in S$ and $f^{-1}((v_i,x_i))=\emptyset$, then we set $f((u_i,v_i)):=(v_i,x_i)$\label{c}.
		\item Otherwise, let $y_i$ be the vertex right after the vertex $x_i$ in $\sigma$. We will show later that $y_i\neq u_i$. Since $(v_i,y_i,u_i)$ is a cycle, there is an arc $a$ in $S \cap \{(v_i,y_i),(y_i,u_i)\}$.  We set $f((u_i,v_i)):=a$\label{d}.
	\end{enumerate}
	%
	%Note that if we enter in case (d), we necessarily have $x_i=u_{i-1}$.
	We now show the correctness of case (d).
	We have to show that $y_i \neq u_i$. Toward a contradiction, suppose that $y_i= u_i$. As we are not in cases (a), (b) or (c), $(v_i,x_i) \in S$ and there exists $(u_j,v_j) \in B$ such that  $f(u_j,v_j)=(v_i,x_i)$ and $j < i$. The arc $(u_j,v_j)$ has been assigned to $(v_i,x_i)$ as a case (b) or (d), thus $u_j=x_i$.
	% We show that $j=i-1$.
	There is at most one vertex between $v_j$ and $v_i$ since otherwise, $v_i \not\in \{x_j,y_j\}$ and thus we would not have $f((u_j,v_j))=(v_i,x_i)$. There is at least one vertex between $v_j$ and $v_i$, since otherwise $\langle v_j,v_i,x_i,u_i \rangle$ would be forbidden pattern $\Pi(U_4)$.
	Therefore, there is exactly one vertex $x_j$ between $v_j$ and $v_i$ then, there is a backward arc adjacent to $x_j$ since otherwise we would have $f((u_j,v_j))=(v_j,x_j)$ or $f((u_j,v_j)) = (x_j,v_i)$ from cases (b) or (c).
	This backward arc is leaving $x_j$, since otherwise this backward arc would be assigned after $(u_j,v_j)$ and therefore $(u_j,v_j)$ would be assigned in same way as before.
	Thus, we have $j=i-1$ and $(x_i,v_{i-1})$ has been assigned to $(v_i,x_i)$ as a case (d).  By induction, it exists a pattern $\langle v_\ell,x_\ell=v_{\ell+1},u_{\ell}=x_{\ell+1},y_{\ell+1}=v_{\ell+2},\dots,y_{i-2}=v_{i-1},x_{i-1}=u_{i-2},y_{i-1}=v_i,u_{i-1}=x_i,u_i\rangle $ in $\sigma$ which is a forbidden pattern. Hence, $y_i\neq u_i$.

	We now show the correctness of $f$.
	First, we show that $f((u_i,v_i))\neq \emptyset$ for every arc of $B$. For cases (a) to (c), $f((u_i,v_i))\neq \emptyset$ since $(v_i,x_i,u_i)$ is a cycle.
	In case (d), $(v_i,y_i,u_i)$  is a cycle and $S \cap \{(v_i,y_i),(y_i,u_i)\}\neq \emptyset$. So, $f((u_i,v_i))\neq \emptyset$.

	Further, we show that for every arc $(s,t) \in S$, we have $|f^{-1}((s,t))|\leq 1$.
	If $(s,t)$ has been assigned as a case (a), then $(s,t)$ is a backward arc and $f((s,t))=(s,t)$ and since it is not possible to assign a backward arc to another backward arc than itself, we have  $|f^{-1}((s,t))|=1$.
	Note that if $(s,t)$ is not a backward arc, then for any backward arc $(u_i,v_i)$, such that $f((u_i,v_i))=(s,t)$, we have either $s=v_i$ and $t\in \{x_i,y_i\}$ (cases (c) or (d)) or $s\in \{x_i,y_i\}$ and $t=u_i$ (case (b) or (d)).
	Hence, $(s,t)$ can be assigned by at most two different backward arcs and $s$ is the head of one them and $t$ is the tail of one of them. Suppose that there exists a backward arc $(u_i,v_i)$ such that $t=u_i$ which is assigned to $(s,t)$ and another backward arc $(u_j,v_j)$ such that $s=v_j$ which is also assigned to $(s,t)$. Suppose that $(u_j,v_j)$ is assigned to $(s,t)$ as a case (c), we then have $s=v_j$ and $t=x_j=u_i$. Since $v_i \infor v_j$, $(u_i,v_i)$ is assigned before $(u_j,v_j)$, we have $f^{-1}((s=v_j,t=x_j=u_i))\neq \emptyset$ when $(u_j,v_j)$ is assigned which is a contradiction.
	Now, suppose that $(u_j,v_j)$  is assigned to $(s,t)$ as a case (d), we then have $s=v_j$ and $t=y_j=u_i$.
	As $(u_i,v_i)$ has been assigned to $(v_j,u_i)$, then $v_j$ is either $x_i$ or $y_i$.
	Moreover, there is another backward arc $(u_\ell,v_\ell)$ such that $f(u_\ell,v_\ell) =(v_j,u_\ell=x_j)$ since we are in case (d).
	As before $v_j$ is either $x_\ell$ or $y_\ell$.
	Therefore, there are two cases: either we have the pattern $\langle v_i,v_\ell,v_j\rangle$ either we have the pattern $\langle v_\ell,v_i,v_j\rangle$.
	In the first case, $(u_i,v_i)$ is assigned to $(v_j=y_i,t=u_i)$ as a case (d).
	Thus, there exists a backward arc leaving $x_i = v_\ell$ which contradicts that $\sigma$ is sparse.
	In the second case, $(u_\ell,v_\ell)$ is assigned to $(y_\ell = v_j,u_\ell)$ as a case (d).
	Hence, there exists a backward arc leaving $x_\ell = v_i$ which contradicts that $\sigma$ is sparse.
	We can conclude that $f$ is an injective function which implies that $|B|\leq |S|$. Hence, $|B|$ is a minimum feedback arc set.

	Finally, since $\sigma$ can be computed in time $O(n^3)$ by \Cref{lemma:no_forbidden_configuration}, a solution of FAST for $T$ can also be computed in polynomial time by taking the backward arcs of $T$ in $\sigma$.
\end{proof}

	\begin{sketch}
		We consider a sparse ordering $\sigma$ of $T$ without forbidden patterns using  \Cref{lemma:no_forbidden_configuration}, as well as any fas $S$ of $T$. Then, to prove that the set $B$ of backward arcs defined by the ordering $\sigma$ is a minimal fas, we construct a function $f : B \to S$ such that, for any $(u_i, v_i)\in B$:
%		\begin{inparaenum}[(a)]
		\begin{enumerate}[(a)]
			\item If $(u_i,v_i)\in S$, then we set $f((u_i,v_i)):=(u_i,v_i)$;
			\item If $(u_i,v_i) \not\in S$ and $(x_i,u_i)\in S$, then we set $f((u_i,v_i)):=(x_i,u_i)$;
			\item If $(u_i,v_i)\not\in S$, $(x_i,u_i)\not\in S$ and $f^{-1}((v_i,x_i))=\emptyset$, then we set $f((u_i,v_i)):=(v_i,x_i)$;
			\item Otherwise, let $y_i$ be the vertex right after the vertex $x_i$ in $\sigma$. We show that if the tournament is without any forbidden pattern, then that $y_i\neq u_i$. Since $(v_i,y_i,u_i)$ is a cycle, there is an arc $a$ in $S \cap \{(v_i,y_i),(y_i,u_i)\}$.  We set $f((u_i,v_i)):=a$.
		\end{enumerate}
%	\end{inparaenum}
		Then we prove that this function injective, proving that $|B| \leq |S|$ and so $B$ is a minimal feedback arc set of $T$.
	\end{sketch}

%%% Local Variables:
%%% mode: latex
%%% TeX-master: "main"
%%% End:

\subsection{FVST is NP-complete on sparse tournaments}
\begin{construction}
	\label{const:FVST to dw}
	Let $G$ be a cubic graph with vertices $\{v_1,\dots,v_n\}$. We construct the following tournament $T$ along with the sparse ordering $\sigma$.
	\begin{compactitem}
		\item For every vertex $v_i$, let $N(v_i) = \{v_j,v_k,v_\ell\}$ be the neighbours of $v_i$ in $G$. We introduce the pattern $p_i = <h_i,u^j_i,u^k_i,u^{\ell}_i,t_i,x^1_i,x^2_i,x^3_i>$.
		\item For every pair of vertices $v_i$ and $v_j$ such that $i<j$, we order $\sigma$ such that $p_i \infor p_j$.
		\item Introduce the following backward arcs. For each vertex $v_i$, construct the backward arc $(t_i,h_i)$ ({\normalfont vertex backward arc}). For every edge $v_i,v_j$ such that $i < j$, construct the backward arc $(u^i_j,u^j_i)$ ({\normalfont  backward arc}). Any other arc is a forward arc.
	\end{compactitem}
\end{construction}

Let $T$ be a tournament and $X$ be a solution for {\sc FVST}.
A backward arc $(t,h)$ is say \textit{saturated} by  $X$ (or simply saturated) if for every vertex $h \infor x \infor  t$, we have $x \in X$. Note that if a backward arc $(t,h)$ is saturated then every cycle $C$ that contains only $(t,h)$ as a backward arc is removed by the $x$ vertices. Similarly, if $(t,h)$ is not saturated, then $\{t,h\} \cap X \neq \emptyset$.

\begin{lemmarep}
	\label{lemma:nice}
	Let $T$ be a tournament resulting from \Cref{const:FVST to dw} along with the sparse ordering $\sigma$. Let $X$ be a solution for FVST in $T$. There is a solution $X'$ such that $|X'|\leq |X|$:
	\begin{compactitem}
		\item for every edge backward arc $(u^i_j,u^j_i)$, we have   $|\{u^i_i,u^j_i\} \cap X'| = 1$, and
		\item for every $v \in X'$, $v$ is adjacent to a backward arc.
	\end{compactitem}
\end{lemmarep}

\begin{proof}
	First, we show that we can construct a solution $X'$ such that for every edge backward arc $(u^j_i,u^i_j)$ we have $\{u^j_i,u^i_j\} \cap X' \neq \emptyset$. Let $(u^i_j,u^j_i)$ be the leftmost edge backward arc such that $\{u^i_j,u^j_i\} \cap X = \emptyset$. It means that $(u^i_j,u^j_i)$ is saturated and so $\{x^1_i,x^2_i,x^3_i\} \subset X$. Let $v_k$ and $v_\ell$ be the two neighbours of $v_i$ different of $v_j$ in $G$. We set $X' = X \cup \{u^j_i,u^k_i,u^\ell_i\} \setminus \{x^1_i,x^2_i,x^3_i\}$. We now show that $X'$ is a solution to FVST. Let $C$ be a cycle containing $x \in \{x^1_i,x^2_i,x^3_i\}$. $C$ necessarily contains a backward arc $(u,v)$ such that $v \infor x \infor u$ and by construction $(u,v)$ is a edge backward arc. If $v \infor u^i_j$ then by hypothesis $\{u,v\} \cap X \neq \emptyset$ and thus, $\{u,v\} \cap X' \neq \emptyset$ which implies that $X'$ removes $C$. Otherwise, we have $v \in \{u^j_i,u^k_i,u^\ell_i\} \subset X'$ and $X'$ removes $C$. Hence $C$ is removed by $X'$. We apply this strategy until there is no edge backward without a vertex in $X'$.
	Further, let $(u^i_j,u^j_i)$ be an edge backward arc such that $\{u^i_j,u^j_i\} \subset X$. We set $X' = X \cup \{h_i\} \setminus \{u^j_i\} $. Let $C$ be cycle containing $u^j_i$. If $C$ contains the vertex backward arc $(t_i,h_i)$ then $C$ is removed by the deletion of $h_i$. Otherwise, $C$ contains an edge backward arc and since every edge backward arc contains a vertex in $X'$, then $C$ is removed by $X'$.\par
	Let $v$ be a vertex in $X'$ such that $v$ is not adjacent to a backward arc. By construction $v$ is $u^j_i$ vertex and thus, any cycle $C$ containing $v$ also contains an edge backward arc $a$. Since every edge backward arc contains a vertex in $X'$ then $X' \setminus \{v\}$ contains the vertex in $a \cap X'$ and thus $X' \setminus \{v\}$ removes $C$. Hence, we can remove $v$ from $X'$.
\end{proof}

%
%	\begin{proofsketch}
%          Suppose $X$ does not respect lemma's properties. 
%          Let $(u^i_j,u^j_i)$ be an edge backward arc. If $|{u^i_j,u^j_i} \cap X | = 0$, then $(u^i_j,u^j_i)$ is saturated, that is, $\{x^1_i,x^2_i,x^3_i\} \subset X$. Let $v_k$ and $v_\ell$ be the two neighbours of $v_i$ different of $v_j$ in $G$. We set $X' = X \cup \{u^j_i,u^k_i,u^\ell_i\} \setminus \{x^1_i,x^2_i,x^3_i\}$. If $|{u^i_j,u^j_i} \cap X | = 2$, then we set $X' = X \cup \{h_i\} \setminus \{u^j_i\} $. Finally, if there is a vertex $v$ such that $v$ is not adjacent to a backward arc, then we set $X' = X \setminus \{v\}$.
%          We then show that in any case, $X'$ is a solution for FVST and we repeat the process until $X'$ respects lemma's properties.
%	\end{proofsketch}

\begin{theoremrep}
	FVST in NP-complete on sparse tournament.
\end{theoremrep}

\begin{proof}
	Let $G$ be a cubic graph and $T$ be a sparse tournament resulting from \Cref{const:FVST to dw} along with the sparse ordering $\sigma$. We show that $G$ contains a vertex cover of size $c$ if and only if $T$ has a solution for FVST of size $c + |E(G)|$.

	Let $S$ be a vertex cover of size $c$ for $G$. We construct a solution $X$ for FVST in $T$. For each vertex $v_i \in S$, we set $h_i \in X$. For each vertex $v_i \not\in S$, let $v_j,v_k$ and $v_\ell$ be the three neighbours of $v_i$. We set $\{u^j_i,u^k_i,u^\ell_i\} \subset X$. Finally, for any edge $v_iv_j$ such that $\{v_i,v_j\}\subset S$, we set $u^i_j \in X$. Let $C$ be a cycle of $T$. If $C$ contains only one backward arc that is a vertex backward arc $(t_i,h_i)$, then either $v_i \in S$ and $C$ is removed by the deletion of $h_i$ or $v_i\not\in S$ and $C$ is removed by the deletions of $u^j_i$, $u^k_i$  and $u^\ell_i$ (where $v_j,v_k$ and $v_\ell$ are the neighbours of $v_i$). Otherwise, $C$ contains an edge backward arc $(u^j_i,u^i_j)$ and since $\{u^j_i,u^i_j\}\cap X \neq \emptyset$, $C$ is removed by $X$. Hence $X$ is a solution for FVST in $T$ and we have $|X| = c+|E(G)|$. \par

	Let $X$ be a solution of size $c+|E(G)|$ for FVST with respect with \Cref{lemma:nice} property. We construct a vertex cover $S$ for $G$. For each vertex backward arc $(t_i,h_i)$, if $(t_i,h_i)$ is not saturated then we set $v_i \in S_i$. Let $v_iv_j$ be an edge of $G$. Since $(u^i_j,u^j_i)$ contains exactly one vertex in $X$, then at least one vertex backward arc among $(t_i,h_i)$ and $(t_j,h_j)$ is not saturated. Thus, either $v_i$ or $v_j$ belongs to $S$ and $v_iv_j$ is covered. Hence, we construct a vertex cover for $G$ of size $c$.
\end{proof}

	\begin{sketch}
          Let $G$ be a cubic graph and $T$ be a sparse tournament resulting from \Cref{const:FVST to dw}. We show that $G$ contains a vertex cover of size $c$ if and only if $T$ has a solution for FVST of size $c + |E(G)|$. Given a vertex cover $S$ for $G$, we can construct a fvs $X$ in $T$ by saturating every vertex backward arc $(h_i,t_i)$ such that $v_i \not\in S$. Then, for any vertex $v_i \in S$, we set $h_i \in X$. Finally, for any edge $v_iv_j$ such that $\{v_i,v_j\}\subset S$, we set $u^i_j \in X$. We then can show that $X$ is a solution for FVST of size $|S| + |E(G)|$.
          Given a solution $X$ of size $c + |E(G)|$ for FVST respecting \Cref{lemma:nice} property, we can construct a vertex cover $S$ by selecting every vertex $v_i$ such that the vertex backward arc $(h_i,t_i)$ is not saturated in $X$. 
	\end{sketch}

%%% Local Variables:
%%% mode: latex
%%% TeX-master: "main"
%%% End:

\section{Conclusion}
In this paper, we studied a new parameter for tournaments, called degreewidth.
We showed that it is NP-hard to decide if degreewidth is at most $k$, for some natural number $k$ and we proceeded to design a $3$-approximation for the degreewidth. One may ask if there is a PTAS for this problem.
Then, we investigated sparse tournaments, \textit{i.e.}, tournaments with degreewidth one and developed a polynomial-time algorithm to compute a sparse ordering. Is it possible to generalise this result by providing an FPT algorithm to compute the degreewidth?
We also showed that FAST can be solved in polynomial-time in sparse tournaments, matching with the  known result that {\sc Arc-Disjoint Triangles Packing} and {\sc Arc-Disjoint Cycle Packing} are both polynomial in sparse tournaments~\cite{BessyBKSSTZ19}.
Therefore, the question arise: can this parameter be used to provide an FPT algorithm for FAST in the general case?
% Second, are there any other problems that can be solved in polynomial time in sparse tournaments? For example, can {\sc Feedback Vertex Set} be solved in polynomial-time in sparse tournament, or will it be NP-hard like {\sc Vertex-Disjoint Triangles Packing}~\cite{BessyBT17}?
%
Furthermore, we showed FPT for DS w.r.t degreewidth. Are there other domination problems e.g., perfect code, partial dominating set, or connected dominating set that is FPT w.r.t degreewidth?
Lastly, we also can wonder if this parameter is useful for general digraphs.
%%% Local Variables:
%%% mode: latex
%%% TeX-master: "main"
%%% End:

%%
%% Bibliography
%%

%% Please use bibtex, 
\newpage
\bibliography{bilbio}

\end{document}